\PassOptionsToPackage{square,comma, numbers,sort&compress}{natbib}  

\documentclass[final,3p,10pt]{elsarticle}
\usepackage[english]{babel}

\newif\ifdraft
\drafttrue

\usepackage{epstopdf}
\usepackage[caption=false]{subfig}
\usepackage{comment}
\usepackage{graphics}
\usepackage{placeins}
\usepackage{textcomp}
\usepackage{color}
\usepackage{amsthm}
\usepackage[normalem]{ulem}
\usepackage{dirtytalk}
\usepackage{csquotes}
\usepackage{url}
\usepackage{marginnote}
\usepackage[htt]{hyphenat}

\usepackage{graphicx}

\usepackage{rotating}

\usepackage{amsmath}  
\usepackage{amsfonts}
\usepackage{amssymb}

\usepackage{url}
\usepackage{multirow}
\usepackage{acronym}
\usepackage[algoruled]{algorithm2e}

\usepackage{color}

\usepackage{ifpdf}

\usepackage{colortbl}
\usepackage{lscape} 
\usepackage{pdflscape} 
\usepackage{array,booktabs,ragged2e}
\usepackage{varwidth}
\usepackage{capt-of}

\theoremstyle{plain}
\newtheorem{theorem}{Theorem}[section]

\theoremstyle{definition}
\newtheorem{definition}[theorem]{Definition}

\theoremstyle{remark}

\definecolor{orange}{RGB}{255,127,0}

\definecolor{gms}{rgb}{0.55, 0.71, 0.0}

\newcommand{\afabcomment}[1]{\ifdraft{\leavevmode\color{red}{[AF]: {#1}}}\else{\vspace{0ex}}\fi}
\newcommand{\gsuscomment}[1]{\ifdraft{\leavevmode\color{blue}{[GSu]: {#1}}}\else{\vspace{0ex}}\fi}

\journal{Information Processing \& Management}

\begin{document}

\begin{frontmatter}

\title{Gender Stereotype Reinforcement:\\ Measuring the Gender Bias Conveyed by Ranking Algorithms} 
\author{Alessandro Fabris}
\ead{fabrisal@dei.unipd.it}

\author{Alberto Purpura}
\ead{purpuraa@dei.unipd.it}

\author{Gianmaria Silvello}
\ead{silvello@dei.unipd.it}

\author{Gian Antonio Susto}
\ead{gianantonio.susto@dei.unipd.it}

\address{Department of Information Engineering, University of Padua, Italy.}

\begin{abstract}
Search Engines (SE) have been shown to perpetuate well-known gender stereotypes identified in psychology literature and to influence users accordingly. Similar biases were found encoded in Word Embeddings (WEs) learned from large online corpora. In this context, we propose the \emph{Gender Stereotype Reinforcement} (GSR) measure, which quantifies the tendency of a SE to support gender stereotypes, leveraging gender-related information encoded in WEs.

Through the critical lens of construct validity, we validate the proposed measure on synthetic and real collections. Subsequently, we use GSR to compare widely-used Information Retrieval ranking algorithms, including lexical, semantic, and neural models. We check if and how ranking algorithms based on WEs inherit the biases of the underlying embeddings. We also consider the most common debiasing approaches for WEs proposed in the literature and test their impact in terms of GSR and common performance measures. 
To the best of our knowledge, GSR is the first specifically tailored measure for IR, capable of quantifying representational harms.

\end{abstract}


\end{frontmatter}






\section{Introduction}
\label{sec:intro}

In a world with zettabytes of data, SEs become the gatekeepers of information. The continuous growth of internet-based content, the maturity of Information Retrieval (IR - the scientific field underlying SEs), along with seamless user experience, contribute to their widespread use. Since the early 2000s, SEs have been utilized by over $90\%$ of internet users\footnote{\url{https://www.pewresearch.org/internet/2012/03/09/search-engine-use-2012/}} and, in the past decade, they have consistently been reported as the most trusted source for general news and information\footnote{\url{https://www.edelman.com/sites/g/files/aatuss191/files/2019-03/2019\_Edelman\_Trust\_Barometer\_Global\_Report.pdf}}. 
Constant availability of information has shaped expectations and cognitive processes of SE users \citep{sparrow2011:ge}. Such factors concur to the importance and relevance of SEs in acquiring knowledge and culture, including perceptions about gender and stereotypes \citep{kay2015:ur}.

Stereotypes can be modelled as associative networks of concepts \citep{sayans2019:is}. They may arise from co-occurrence of features \citep{lepellley2010:sf}, such as membership to a group and display of certain traits and roles, which become linked in a Bayesian fashion based on culture and direct observation \citep{hinton2017:is}. Women and men are particularly salient categories, recognizable since an early age, and available for stereotypical association with traits, behaviors and events  \citep{martin2010:pg}. In turn, even when outspokenly rejected, gender stereotypes influence the lives of women and men both descriptively and prescriptively, shaping the qualities, priorities and  needs that members of each gender are expected to possess \citep{ellemers2018:gs}. During their lives, individuals are frequently exposed to information about gender, through direct experience and indirect information coming from social interactions and cultural representations \citep{eagly2019:gs}, often portrayed by the media. 

Cultivation theory \citep{gerbner1986:lt}, historically focused on television, posits that increasing exposure to a medium and its contents leads to a progressive alignment to the beliefs, culture and reality depicted in the televised world. Within this framework, the way women are depicted in primetime television has been studied; recent analysis highlights persistent representational stereotypes related to physical appearance and warmth \citep{sink2017:dg}, confirmed by public opinion \citep{eagly2019:gs, tantleff2011wt}. According to cultivation theory, heavy viewers are likely to be influenced in their perception of the real world, due to the availability heuristic \citep{shrum1995:as}: in judging frequency and normality (e.g.\ of women being affectionate), they resort to the examples that come to their mind, the media being a potential source of information to recall. The availability heuristic has been proposed and verified as a general shortcut in human cognitive processes \citep{tversky1973:ah}, and recently studied as a bias that arises while exploring result pages from SEs \citep{novin2017:ms}.


Inevitably, SEs influence users, helping them to link topics, concepts and people as they read, browse and acquire knowledge. They therefore can play an important role in countering or reinforcing stereotypes. For instance, search results on Google images were found to reflect current gender differences in occupation, with a tendency to slight exaggeration \citep{kay2015:ur}; at the time of the study, searching images of a job with a female-to-male ratio of 1:4 in the employed population, such as software engineer, would yield pictures depicting women in less than $20\%$ of the results. Moreover, manipulation of female-male representation in search results about a job, artificially increasing the presence of one gender in images, significantly impacted people's perception about gender ratios in that occupation \citep{kay2015:ur}.  A study on Bing photos found a greater frequency of women in depictions of warm traits (e.g.\ sensitive), while men are more common in searches about competence traits (e.g.\ intelligent) \citep{otterbacher2017:cm}. These results highlight the importance of measuring and countering bias in SEs, as recently pointed out by critical race and gender studies scholarship \citep{noble2018:ao}. 

Gender stereotypes held by people are commmonly measured in two ways: directly, on the basis of in individual agreeing with statements about gender and specific traits \citep{eagly2019:gs}; indirectly, via Implicit Association Tests (IAT) between mental representations of objects \citep{greenwald1998:mi} or assessment of attitude through priming \citep{fazio1995:va}. Indirect tests are appealing as they allow an unobtrusive assessment of attitudes towards groups (determined e.g. by gender and ethnicity) and can measure association of categories, such as \texttt{women}, with words from a specific domain, such as \texttt{family}, even when subconscious. Large text corpora sourced from the web, such as Wikipedia, have been found to echo some of the above biases: as an example, Wikipedia entries related to women are more likely to mention marriage- and sex-related contents and events \citep{graells2015:fw}. Interestingly, Word Embeddings can be used to detect gender-related biases in the corpus they have been trained on \citep{garg2018:we, defranza2020:ls}. 

Word Embeddings (WEs) are vectorial representations of words computed automatically using different supervised and unsupervised machine learning approaches \citep{joulin2016bag, mikolov2013:dr}. Most frequently, they are learnt from large text corpora available online (such as Wikipedia, Google News and Common Crawl, capturing semantic relationships of words based on their usage. Recent work \citep{caliskan2017:sd} shows that WEs retain the stereotypical associations from their training corpora, encoding a full spectrum of biases from the IAT, including gender-related ones about career and family, science and arts. Additional problematic depictions of men and women have been identified in these WEs, including sexist analogies (such as $\texttt{woman}-\texttt{man} \simeq \texttt{midwife} - \texttt{doctor} \simeq \texttt{whore} - \texttt{coward}$ \citep{bolukbasi2016:mc})
 and representation of jobs skewed with respect to gender, in ways that reflect current gender gaps in the US workforce \citep{garg2018:we, dearteaga2019:bb, prost2019:de}. For this reason, WEs have been proposed as an unobtrusive measurement tool of the average bias of the many contributors to these corpora and, generalizing, from the society they live in \citep{garg2018:we} or the language they speak \citep{defranza2020:ls}. Based on co-occurrence with intrinsically gendered terms within the text corpora (such as \texttt{woman} and \texttt{man}), a \emph{genderedness} score can be derived for each word in the embedding space. Among words with a high score, some are duly gendered (\texttt{hers}, \texttt{his}), while others reflect an accidental status quo aligned with stereotype (\texttt{hygienist}, \texttt{electrician} - stereotypically female and male, respectively). This bias, undesirable when WEs are part of a socio-technical system, is an interesting property we can leverage to measure gender stereotypes in SEs.

In this work, we propose the \emph{Gender Stereotype Reinforcement} (GSR) measure that is specifically tailored to quantify the tendency of a SE's ranked list to support gender stereotypes. GSR exploits gender bias encoded in WEs to detect and quantify the extent to which a SE responds to stereotypically gendered queries with documents containing stereotypical language of same polarity. 

Firstly, we validate the word-level genderedness score against well-studied gender stereotypes and subsequently, we employ a basic compositional model to quantify whether retrieved documents are connected to queries along stereotypical lines. We operazionalize GSR based on this model, and verify its ability to capture \emph{direct} and \emph{indirect gender stereotype} reinforcement on synthetic and real collections.

Secondly, we audit IR ranking algorithms from different families: (i) lexical models, including \texttt{BM25} \citep{RobertsonZaragoza2009}, Query Likelihood Model (\texttt{QLM} - \citep{Zhai2008}) and \texttt{tf-idf} \citep{SaltonMcGill1983}; (ii) semantic models, such as those using Word2Vec with additional compositionality  (\texttt{w2v\_add}) and with self-information (\texttt{w2v\_si}) \citep{vulic2015:mc}; (iii) neural architectures, including Deep Relevance Matching Model (\texttt{DRMM} - \citep{Guo2016}), and Match Pyramid (\texttt{MP} - \citep{pang2016:tm}).  We measure each system's performance and GSR on the Text REtrieval Conference (TREC)\footnote{http://trec.nist.gov/} Robust04 \citep{harman1992:tp}  curated and widely-used news-based collection. We also perform qualitative analysis of queries with the highest genderedness score and we find these queries to mirror gender stereotypes studied in psychology literature. 
Moreover, we analyze these ranking models to investigate whether semantic and neural models inherit problematic gender associations from underlying WEs, while verifying the neutrality of lexical models in this regard. We also investigate the tradeoff between performance and fairness for these families of models.

Thirdly, we assess the impact of debiasing WEs \citep{bolukbasi2016:mc} in the context of IR, confirming recent findings that gender-related information is redundantly encoded along multiple directions \citep{gonen2019:lp}. We conclude that the genderedness score, estimated by the proposed GSR measure, is a good proxy for gender bias \citep{garg2018:we}; and, that treating WEs to neutralize it is not sufficient to enforce a real lack of gender bias in word representations and downstream tasks.

Finally, we discuss the construct validity and reliability of our measurement model \citep{messick1998:tv,jacobs2019:mf}. We decouple GSR as a \emph{construct} (the unobservable theoretical abstraction we aim to characterize), from its \emph{operationalization} (how we measure it),  and elucidate the underlying assumptions and properties it should capture. In such context, we argue that some clustering of language along a gendered dimension captured by WEs is inevitable due to domain-specificity of language. The reliability of GSR is evaluated by testing its stability when computed based on WEs learned from different corpora and learning architectures, ensuring that the measurement is robust and not overly dependent on choices of training set and architecture.\\

\noindent \textbf{Contributions} of this work include: 
\begin{enumerate}
    \item GSR measure tailored for SEs and its evaluation within the \emph{construct validity} framework;
    \item audit, in terms of GSR, of several widely-known and used ranking algorithms;
    \item estimation of the impact of different WE debiasing approaches, both on ranking effectiveness and countering gender bias.
\end{enumerate}

\noindent \textbf{Outline}. The rest of this paper is organized as follows. Section \ref{sec:related} describes related works from different domains as IR, Natural Language processing (NLP), algorithmic auditing, social psychology and validity theory. The GSR measure is described in Section \ref{sec:proposed}, preceded by a detailed definition of the abstract construct we aim to quantify, and followed by a toy example that favors a discussion of its key properties. Experiments on real and synthetic IR collections are reported in Section \ref{sec:experiments},  while Section \ref{sec:conclusion} summarizes our conclusions and outlines future works.

\section{Related work}
\label{sec:related}

\subsection{WEs and neural models in IR}
\label{sec:we_ir}
Word2Vec \citep{mikolov2013:dr} was the first widely used WE model. 
Word2Vec can learn similar representations for terms used in similar contexts in the training data, typically corpora of millions of documents in natural language. In addition, 
as the embedded word representations learned with Word2Vec reflect the usage distribution of respective terms, they have been employed as a proxy for the semantic similarity of terms in many NLP applications. The popularity of Word2Vec also paved the way to other machine learning approaches to obtain embedded word representations such as GloVe \citep{pennington2014glove} and FastText \citep{joulin2016bag}.
WE models were soon adopted in the IR domain, promoting the exploration of deep learning approaches for document retrieval~\citep{mitra2018introduction}.

Lexical approaches such as \texttt{tf-idf} \citep{SaltonMcGill1983}, \texttt{QLM} \citep{Zhai2008} and \texttt{BM25} \citep{RobertsonZaragoza2009} were the first and most popular techniques adopted for document retrieval. 
Nevertheless, these retrieval models do not take into account terms which are not contained in the user query nor their semantics. For this reason, the potential offered by embedded word representations -- i.e.\ the possibility to represent the meaning of a term and compare it to others in a measurable way -- was soon put to use by newly proposed retrieval models.

The simplest WE-based document retrieval approach in our experiments is named \texttt{w2v-add} \citep{vulic2015:mc}. In this case, we compute a query and a document representation averaging the WEs of the terms they contain, and then rank documents according to their cosine similarity to the query vector. This approach however reduces the query/document representation problem to the core. For example, it does not take into account the relative importance of each term. \texttt{w2v-si} solves this problem: queries and documents representations are obtained computing a weighted average of the word vectors of their terms and then documents are ranked as in the previous case. Each term weight corresponds to its self-information (si) which is a term specificity measure similar to IDF \citep{cover2012elements}.

Among the first most successful deep learning models for IR, there is Deep Relevance Matching Model (\texttt{DRMM} - \citep{Guo2016}). \texttt{DRMM} uses embedded representation of words to compute the similarity between every pair of terms in a user query and each document in a ranked list. 
Another paradigmatic approach in the Neural IR field is MatchPyramid (\texttt{MP} - \citep{pang2016:tm}). This approach, originally proposed as a document classification model, was also successfully applied to the ranked task. 
For our study, we select these two approaches, being popular in IR and easy to use. Moreover, they allow us to evaluate the impact of diverse WEs in different Neural IR architectures.



\subsection{Gender stereotype in WEs and SEs} \label{sec:Gender stereotype in WEs and SEs}

A convincing body of research shows that WEs learnt on large corpora of text available online encode cultural aspects, some of which undesirable. Among them, worth noting at the core of this work are gender-related biases which comprise: sexist analogies \citep{bolukbasi2016:mc}, stereotypical association of gender with science and arts \citep{caliskan2017:sd}, representation of occupations correlated to differences in female and male employment \citep{garg2018:we, dearteaga2019:bb, prost2019:de}, gender roles in career and within the family \citep{caliskan2017:sd}. Communion (also called warmth) and agency are two further dimensions consistently associated with gender \citep{eagly2019:gs}, analyzed in Section \ref{sec:map1}; in line with this stereotype, we find warm traits (e.g.\ ``emotional'') to have female polarity in the embedding space, while agentic traits (e.g.\ ``aggressive'') are more commonly associated to men  (Section \ref{sec:map1}). A wealth of studies in psychology and labor economics literature confirms the presence of the above-mentioned biases in society \citep{fiske2002:ms, nosek2002:hi, nosek2002:mm, cvencek2011:mg, eagly2019:gs, hentschel2019:md, bls2019}, makeing their presence in WEs particularly interesting.

Several of these biases, found in SEs, potentially reinforce gender stereotype through powerful and pervasive search tools available to the public. \citet{kay2015:ur} show that gender bias in image search results is exaggerated: the gender distribution for Google image results about jobs are correlated to and amplify differences in female and male employment. Bing images associate agentic traits to men and warm traits to women \citep{otterbacher2017:cm}. Monster and CareerBuilder were audited, displaying group unfairness against female candidates in 1/3 of the job titles surveyed \citep{chen2018:ii}. This does not imply that these SEs are likely to have the same biased WEs as part of their algorithmic machinery. Rather, finding that known gender biases in SEs are also encoded in vectorial representations of words suggests that WEs can be used as a tool to measure gender bias in SEs.

In this respect, \citet{bolukbasi2016:mc} find that gender-related information for each word in the embedding space is mostly confined within a single dimension:
\begin{enumerate}
    \item They propose ten word pairs to define gender: \texttt{she}-\texttt{he}, \texttt{her}-\texttt{his}, \texttt{woman}-\texttt{man}, \texttt{Mary}-\texttt{John}, \texttt{herself}-\texttt{himself}, \texttt{daughter}-\texttt{son}, \texttt{mother}-\texttt{father}, \texttt{gal}-\texttt{guy}, \texttt{girl}-\texttt{boy}, \texttt{female}-\texttt{male}.
    \item For each pair they compute the difference between the two word vectors, obtaining ten candidate vectors (dimensions) to encode gender.
    \item They stack the ten vectors into a single matrix, on which they perform a principal component analysis, finding 60\% of the variance explained by first principal component, subsequently treated as the gender subspace $w_g$. We dub \emph{genderedness} score of a word $w$, its scalar projection along the gender subspace 
    \begin{align}
        \label{eq:gend_score}
        g(w) &= \frac{w \cdot w_g}{|w||w_g|},
    \end{align}
    \noindent and use it as a building block to operationalize GSR (Section \ref{sec:map1}).
\end{enumerate}

\noindent The sign and magnitude of $g(w)$ determine the polarity and strength of gender-association for word $w$ - e.g.\ $g(\texttt{sister})=0.31$, $g(\texttt{brother})=-0.22$. After identifying a gender subspace (or direction $w_g$), \citet{bolukbasi2016:mc} remove gender-related information from most words via orthogonal projection. Only intrinsically gendered word pairs (such as \texttt{she}, \texttt{he}) retain a non-zero component in the gender subspace. \citet{prost2019:de} propose a \emph{strong} variant of this approach where the procedure applies to the whole vocabulary. This family of debiasing techniques seem limited and imperfect \citep{gonen2019:lp}, with gender information redundantly encoded along multiple dimensions, and thus hard to eradicate. Confirmation of this statement is given in the context of SEs and gender stereotype in Section \ref{sec:deb}, where we assess the impact of regular and strong debiasing with respect to performance and GSR of IR models based on WEs.

\subsection{Fairness and diversity in IR} 

Fairness in information retrieval and recommendation is an area of increasing interest for academia and industry, with entire tracks\footnote{\url{https://fair-trec.github.io/}}, workshops\footnote{\url{http://bias.disim.univaq.it/}} and corporate teams devoted to such a complex topic. Efforts in the field are aimed at emphasizing the social context which SEs and recommender systems inhabit and influence.

\noindent We borrow from \citet{ekstrand2019:fd} in sketching a taxonomy of fairness in search along two dimensions: the people benefiting from our efforts and the type of harm we are trying to prevent. Based on their position in the information pipeline, fairness can benefit:
\begin{enumerate}
    \item consumers of documents (e.g.\ SE users), when focused on user privacy \citep{fu2015:ep}, content diversity \citep{gao2020:tc} or targeted advertisement, which may imply unequal opportunity for different segments of the population \citep{celis2019:tc};
    \item providers of documents, who deserve an equal chance to be read, viewed and clicked \citep{yang2017:mf, zehlike2017:ft, singh2018:fe, wu2018:dd};
    \item information subjects (mentioned in documents or queries), who may be present in contexts where they would rather not appear or, conversely, neglected or censored out against their will \citep{bamman2012:cd, kay2015:ur, otterbacher2017:cm, noble2018:ao}.
\end{enumerate}

\noindent Harms can be:
\begin{enumerate}
    \item distributional, when related to a resource of interest, such as education opportunities, jobs, access to credit, possibility of parole \citep{zehlike2017:ft}, exposure \citep{singh2018:fe}, or, more generally, attention of consumers using a SE over time \citep{biega2018:ea}, from which the above-mentioned resources depend when decision-making is not fully automated;
    \item representational, likely to take place when individuals and groups are unable to self-determine their image, which may end up being stereotyped, inadequate or offensive \citep{otterbacher2017:cm, noble2018:ao, abbasi2019:fr}. In the context of SEs, representational harms typically refer to information subjects, and our work is no exception: GSR by SEs is firstly a type of problematic representation of women and men who happen to be information subjects in search and browsing experiences of SE users. 
\end{enumerate}

A subfield of research, often referred to as \emph{fair ranking}, addresses distributional harms for providers \citep{yang2017:mf, biega2018:ea} and consumers \citep{celis2019:tc}. This work is typically aimed at minimizing disparities in the outcomes of similar individuals (\emph{individual fairness}) or groups determined by a protected attribute such as gender, ethnicity, religion (\emph{group fairness}). These approaches perform re-ranking of results which have been retrieved and ranked by a supposedly biased algorithm. An alternative paradigm aims at directly modifying the retrieval algorithm. \citet{gerritse2019:id}\footnote{We refer to the extension of this work discussed in ECIR 2020 workshop on Algorithmic Bias in Search and Recommendation (\url{http://bias.disim.univaq.it/}) whose proceedings are currently unavailable.} studies the impact of debiasing WEs \citep{bolukbasi2016:mc} in algorithms of query reformulation based on Word2Vec embeddings. This work is the closest to our evaluation of the effects of debiasing in Section \ref{sec:deb}, where we perform a complementary analysis on different IR algorithms which are purely based on WEs.

Metrics and approaches from fair ranking can also be employed to measure and favor a diversified topical coverage \citep{gao2020:tc}, where political leaning or sentiment take on the role of a protected attribute which should have reasonable diversification across search results. This flavor of fairness overlaps with diversity and novelty research from the IR community \citep{clarke2008:nd, carpineto2012:es, yu2018:rc}. 

In some areas, such as political search in social media, it is interesting to evaluate how diversity and bias in search results can be influenced by (implicit) bias in queries. For instance, \citet{kulshrestha2017:qs} find that Twitter's response to queries about US political candidates tends to give better ranking to tweets from sources with the same political leaning as the candidate. Although different in methods and objective, our work is conceptually similar as we are interested in evaluating how  a construct measured on queries (stereotypical genderedness) relates to the same construct measured on search results. 

It should be noted that the taxonomy we presented is far from complete. A thorough categorization of ongoing efforts to audit and improve the fairness and trustworthiness of SEs would be as complex as the underlying socio-technical systems. Further considerations may include a temporal dimension and a spectrum to quantify division within a community, as in the case of research on echo chambers and filter bubbles \citep{flaxman2016:fb}. User interfaces also play a key role: responsible augmentation of search results may be important to convey information about fact checking \citep{zhang2020:oo} and controversy \citep{zielinski2018:cc}; query auto-completion can lead to problematic results \citep{noble2018:ao}, while panels which summarize results for users within the SE may reduce click-through rates for the websites from which information is extracted.

\subsection{Construct validity and reliability}
\label{sec:construct_val}
Construct validity, in its modern connotation, is a unified view on the desired properties for a measure aimed at quantifying a given construct that enables an overall judgement about adequacy and appropriateness based on empirical evidence and theoretical rationales \citep{messick1995:vp}. Embedded in this definition is a clear distinction between, on the one hand,  the unobservable theoretical attribute we are trying to evaluate (the \emph{construct}, e.g.\ ``teacher quality''), with its context and underlying theme and, on the other, the way the construct becomes operational through a measurement model (the \emph{operationalization}).

We follow \citet{jacobs2019:mf}, who describe seven components of construct validity, which we summarize below:

\begin{enumerate}
    \item \emph{Face validity}. How plausible does the measurement model look compared to the construct? Answers to this question are highly subjective and little more than a preliminary step.
    \item \emph{Content validity}. Is there a coherent understanding of the theoretical construct? Is the selected operationalization in accordance with it?
    \item \emph{Convergent validity}. Does our measurement agree with other measurements of the same construct?
    \item \emph{Discriminant validity}. What else is the measurement capturing? Are there other constructs which are justifiably or unexpectedly correlated with the proposed measurement? 
    \item \emph{Predictive validity}. Are any other properties likely to be influenced by our construct? Is our operationalization of the construct related to those properties as expected?
    \item \emph{Hypothesis validity}.  Are the construct and its operationalization meaningful and useful, so that they can be used to test hypotheses and raise new  questions?
    \item \emph{Consequential validity}. Should our measure be used? In which context can it be employed and what would the be consequences? 
\end{enumerate}

Section \ref{sec:construct} describes in detail Gender Stereotype Reinforcement (GSR) as a construct, referring to supporting literature from social psychology, which deals with the common understanding of GSR and its \emph{content validity} as a construct. The reliability of our operationalization with reference to the construct is addressed through discussion (Section \ref{sec:operionalization}) and experiments (Section \ref{sec:gsr_dir_stereo}). Considering key properties of GSR, Section \ref{sec:gsr_properties} discusses its \emph{discriminant validity}, tied to domain-specificity of language, along with its \emph{convergent validity} in a wider context of fairness metrics. \emph{Consequential validity} and \emph{hypothesis validity} are linked with current limitations and future work, discussed in Section \ref{sec:conclusion}. In the absence of a user study \emph{predictive validity} cannot be properly discussed. Within the context of gender stereotypes in SE, the only user study we are aware of centers on image retrieval \citep{kay2015:ur}, while our proposed measure deals with textual data. Due to its subjective nature, we do not specifically address \emph{face validity}.

We also discuss GSR \emph{reliability},  a more familiar concept to computer scientists. It depends on stability of measured quantity, precision of measurement tool, and process noise, to determine how robust repeatable and reliable a measure is; Section  \ref{sec:reliability} is devoted to this aspect.

\section{Proposed approach}
\label{sec:proposed}

We articulate our approach, untangling the definition of a \emph{construct}, i.e.\ the phenomenon we want to study, from its subsequent \emph{operationalization}, which details how the phenomenon can be measured from observed data \citep{messick1998:tv, jacobs2019:mf}.

\subsection{Construct}
\label{sec:construct}

Our aim is to quantify to what extent a SE can reinforce gender stereotypes in users. We call this construct \emph{Gender Stereotype Reinforcement} (GSR), resorting to supporting concepts from psychology literature before giving a formal definition. This incremental process is important to establish the \emph{content validity} of GSR as a construct.

\begin{definition}{Stereotype}
  
\noindent Stereotypes are beliefs about groups of individuals with a common trait,  widely held by a population of interest. Their appearance is likely influenced by the strength of an observational link, i.e.\ how often one position along a dimension (such as gender) co-occurs with another (such as warmth)
\citep{lepellley2010:sf}.
\end{definition}

Stereotypical associations picked up by individuals can be attributed to culture and socialization \citep{hinton2017:is}. Bayesian principles are thought to be at play in the acquisition of culture, which is often screened and mediated by search technology, whose trustworthiness is generally taken for granted \citep{halavais2008:se}. In other words, our cognition is receptive to repeated co-occurrence of topics and entities. It may therefore end up forming links between them, also thanks to the media and technology we interact with on a daily basis.

\begin{definition}{Gender Stereotype}

\noindent A gender stereotype is a generalised view or preconception about attributes or characteristics, or the roles that are or ought to be possessed by, or performed by, women and men. \footnote{https://www.ohchr.org/en/issues/women/wrgs/pages/genderstereotypes.aspx}
\end{definition}

Stereotypes about gender have been studied in a variety of contexts, including school \citep{cvencek2011:mg}, workplace \citep{bobbitt2011:gd}, parenthood \citep{cuddy2004:pb} and search for romantic partners \citep{park2015:pd}, with respect to several aspects such as depiction, perception (of self and others) and outcomes. 
Common themes have been identified through decades of scholarship, including \emph{agency} and propensity to science, \emph{communion} and importance of appearance~\citep{ellemers2018:gs}. 

As a well-researched example, historical meta-analysis over seven decades confirms agency and communion as consistently and increasingly salient in U.S. opinion polls about gender differences  \citep{eagly2019:gs}. Agency, perceived as predominantly male, refers to drive for achievement, while communion is related to caring for others and is increasingly associated to women.

\begin{definition}{Direct gender stereotype}
\label{def:dgs}
  
\noindent Association of a stereotypically gendered concept with people of the respective gender.
\end{definition}

\noindent This applies to any sentence where preconceptions about one gender are directly associated to a member of that gender, mentioned through either a noun (\texttt{man}), adjective (\texttt{his}), pronoun (\texttt{he}) or name (\texttt{John}). 

\noindent Example: \emph{She is affectionate}.

\begin{definition}{Indirect gender stereotype}
\label{def:igs}

\noindent The link of a stereotypically gendered concept with another stereotypically gendered concept, commonly associated to the same gender.
\end{definition}

\noindent This definition is based on a view of culture, social constructs and stereotypes as networks of concepts \citep{patterson2007:wk, ghosh2014:ms} and implicit associations \citep{fazio1995:va, greenwald1998:mi, berinsky2005:ie}. Co-occurrence of stereotypical characteristics and traits, commonly associated to one gender, may reinforce a link in a network of stereotypes about women and men. To exemplify, we argue that beliefs about stereotypically female (male) jobs are likely to fall on women (men). Research from social cognition and political science highlights that networks of stereotypes associated with protected attributes, such as gender and ethnicity, can play a role in a person's perception, without them being aware of it \citep{berinsky2005:ie}. This may happen to a person, even if they sincerely dislike said stereotype \citep{fazio1995:va}.

\noindent Example: \emph{The nurse is affectionate}

\vspace{0.2cm}

\noindent \textbf{Characterization of the GSR construct.} 

\noindent Given the above terminology, we characterize GSR in the context of IR as the SE's tendency to respond to stereotypically gendered queries with documents containing stereotypical language with the same polarity. We defer a thorough definition, complete with mathematical formalization, to Definition \ref{def:gsr2}.

\noindent In societal systems, GSR is measured by the agreement of human constituents with gender stereotype descriptors \citep{tresh2019:er, plante2013:gs}. 
In operationalizing this construct, we aim to quantify the impact of SEs on the perception of gender: more specifically, its alignment to existing direct and indirect stereotypes encoded in culture and language. Search results may end up reinforcing gender stereotypes if, when responding to potentially stereotypical queries, their language is skewed along gendered lines with matching polarity.

Intuitively, the influence that people around us may exert can be regarded as the societal counterpart of documents and their language in the context of SEs. An example is a SE which, responding to a query about nursing, displays documents with a strong representation of women (\emph{direct} stereotype), or emphasis on attributes related to communion (\emph{indirect} stereotype).

\subsection{Operationalization}
\label{sec:operionalization}

After defining our construct, we show how it can be made operational. This entails illustrating our assumptions and their interplay with the building blocks of our measurement model \citep{jacobs2019:mf}. We begin by defining the basic concepts in the context of search.

\begin{definition}{Ranked list}

\noindent The response of a SE to a query, i.e.\ a ranked list $\mathcal{L}$ of documents, decreasingly ordered by (estimated) relevance with respect to a given user query.
\end{definition}

\begin{definition}{Search history}
\label{def:sh}

A set of (query, ranked list) pairs representing the interactions of one or more users with a SE.

\end{definition}

Stereotype formation may be conceptualized as an acquisition of culture and associations of a particular kind, taking place through \emph{repeated} interaction. The response to a single query, though anecdotally interesting, is less informative than a set of responses to different queries. Hence, we refer to a \emph{search history}, a somewhat overloaded expression, which potentially encompasses every past user interaction with a SE, including the pages they visited along with very detailed logs of click behavior, browsing and permanence.

Our usage of the expression is different in two ways. (1) It applies to any subset of user interactions with a SE, including for instance only recent ones. We do not require a complete list of queries issued and results shown. (2) The level of granularity and depth of logging entailed by our definition is minimal. This work is aimed at auditing and modeling SEs rather than users. For this reason we do not require click logs, which are user-dependent and thus accidental with respect to our analysis. More in general, Definition \ref{def:sh} adapts to data coming from multiple users in a bundled and anonymized fashion, as well as data collected and curated by a practitioner. These differences are important to correctly assess the applicability, practicality and ethics of our operationalization.

To summarize the following sections, we assume that a strong correlation between \emph{genderedness} of queries and of ranked document list in a search history reinforces gender stereotype. In the following, we gradually introduce related quantities; the adopted notation is summarized in Table \ref{tab:notation}.

\begin{table}[htp]
  \begin{center}
    \begin{tabular}{|r|l|}
      \hline
      $\mathcal{Q}$ & set of queries in search history \\
      $\mathcal{D}$ & set of available documents \\
      $\mathcal{L}$ & ranked list of documents \\
      $N = |\mathcal{Q}|$ & number of queries in search history \\
      $w$ & a word \\
      $q \in \mathcal{Q}$ 	&  a specific query \\
      $d \in \mathcal{D}$ 	&  a specific document \\
      $g(w)$ & genderedness of word $w$ \\
      $g(q)$ 	& 	genderedness of query $q$ \\
      $g_q(d)$ 	& 	genderedness of document $d$ retrieved for $q$ \\
      $g_q(\mathcal{L})$   &    genderedness of ranked list $\mathcal{L}$ retrieved for $q$ \\
      $r_k$ & rank of document $d_k$ in list $\mathcal{L}$ \\
      $\mu_q$ 	& 	average genderedness of queries from $\mathcal{Q}$\\
      $\sigma_{g(q)}^2$ 	& variance in genderedness of queries\\
      $\mu_{q,\mathcal{L}}$   &   average genderedness of ranked lists of documents \\
      $m_{s}(\mathcal{Q}, \mathcal{D})$  &  GSR for system $s$ on collection $(\mathcal{Q}, \mathcal{D})$.\\
      \hline
      \end{tabular}
    \caption{Notation for proposed measure.}
    \label{tab:notation}
    \end{center}
\end{table}

\subsubsection{Measuring gender stereotype}
\label{sec:map1}
Stereotypes about gender are plentiful and pervasive, likely due to the fact that the underlying categories (especially the classical female-male dichotomy) are available to our cognition from an early age on a daily basis. Preferential association of a concept or topic to men or women is measured by surveying a population of individuals. The study of gender-based associations thus depends on resources, time available and research agendas.

\begin{figure}[!tbp]
  \centering
  \includegraphics[width=0.8\textwidth]{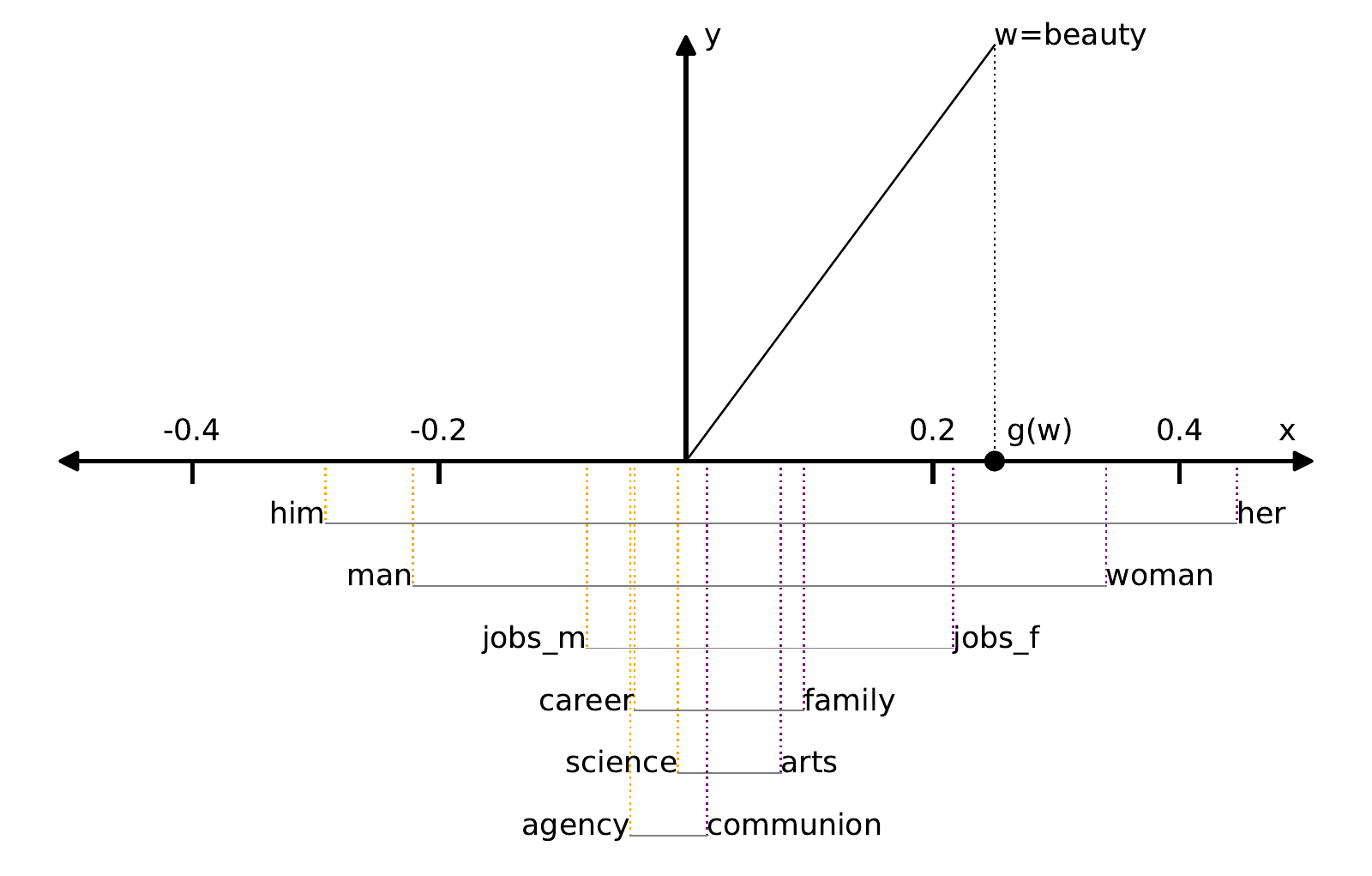}
  \caption{Computation of genderedness for $w=\texttt{beauty}$. Gender direction is on $x$ axis \citep{bolukbasi2016:mc}, while $y$ axis represents subspace orthogonal to gender.  $w$ displays a significant component along the gender direction, hence we consider it stereotypically female in the embedding space and the underlying text corpus. See, in \ref{sec:app},  Table \ref{tab:agency_vs_communion} for \emph{agency} vs \emph{communion} ($p=2.2\mathrm{x}{10^{-2}}$), Table \ref{tab:science_vs_arts} for \emph{science} vs \emph{arts} ($p=1.8\mathrm{x}{10^{-3}}$), Table \ref{tab:career_vs_family} for \emph{career} vs \emph{family} ($p=5.5\mathrm{x}{10^{-4}}$), Table \ref{tab:job} for \emph{jobs\_m} vs \emph{jobs\_f} ($p=0$). P-values are computed with four one-tailed permutation tests on the genderedness of the words in each table.}
  \label{fig:map1}
\end{figure}

Increasing evidence from the field of NLP shows that, among the powerful results and interesting properties of WEs, their geometry  captures well-known stereotypes related to gender \citep{caliskan2017:sd, bolukbasi2016:mc, garg2018:we, prost2019:de, dev2019:ab, papakyriakopoulos2020:bw}. Techniques have been proposed to isolate a word's genderedness along a single direction \citep{bolukbasi2016:mc}. Based on this approach, each word is associated to a ``gender score'' consisting of a signed scalar value. In a convention employed hereafter, a strongly positive (negative) score will be a proxy for a strong association to female (male) gender. The upper part of Figure \ref{fig:map1} depicts, as a simplified example, the projection of the word \texttt{beauty}, which is strongly positive and thus associated to female gender. \footnote{For obvious reasons, a figure can only represent 2 out of the 300 dimensions in which the \texttt{w2v} embedding is encoded.}

To validate genderedness, encoded by Equation \ref{eq:gend_score}, as a score of perceived masculinity/femininity, we test it against known gender stereotypes. Two commonly studied constructs in psychology literature are \emph{agency} and \emph{communion} \citep{eagly2019:gs, hentschel2019:md}, alternatively dichotomized as \emph{competence} and \emph{warmth}  \citep{fiske2002:ms}. Agency, stereotypically associated to men, is related to ability and drive to pursue one's goals, while displaying leadership and assertiveness. Communion, prevalent in female stereotypes, relates to a person's orientation towards others and their well-being, suggesting propensity for caring, nurturing, compassion and emotion. 

Attitude towards mathematics and sciences have been measured implicitly  \citep{cvencek2011:mg, nosek2002:mm} and explicitly  \citep{cvencek2011:mg}. Studies provide evidence of cognitive link between math and male gender from an early age. This association is often studied in opposition to arts (and language) which are found to be predominantly associated with female gender \citep{nosek2002:hi, nosek2002:mm}.

Career orientation, in opposition to family, is another dimension related to gender \citep{nosek2002:hi}. Career can also be broken down into sector. Some professions have a very high male representation, while other work is overwhelmingly carried out by women \citep{bls2019}.

In considering research on gender stereotypes four opposing associations emerged, which are described above. We compute their genderedness as follows: for \emph{agency} vs \emph{communion} we summarize the genderedness of either construct with the average genderedness of adjectives in Table \ref{tab:agency_vs_communion}, taken from \citep{eagly2019:gs}. With the same averaging procedure, we follow \citep{nosek2002:hi} for terms related to \emph{science} vs \emph{arts} (Table \ref{tab:science_vs_arts}) and \citep{nosek2002:mm} for \emph{career} vs \emph{family} (Table \ref{tab:career_vs_family}). Finally, we sample the 20 most gendered single-word jobs from \citep{bls2019}, shown in Table \ref{tab:job}, and perform the same computation, dubbing this comparison \emph{jobs\_m} vs \emph{jobs\_f}. 

Results are summarized in the lower part of Figure \ref{fig:map1}, where we also report the projections of \texttt{woman}, \texttt{man}, \texttt{her}, \texttt{his} for comparison. All four stereotypes are confirmed, with male clusters' projections (orange) falling to the left of their female counterparts (purple). According to one-tailed permutation tests, the dichotomy \emph{agency} vs \emph{communion} is the least gendered,  significant at $p=2.2\mathrm{x}{10^{-2}}$. Interestingly, the strongest association with gender is \emph{jobs\_m} vs \emph{jobs\_f} ($p=0$), stemming from census data and representing occupations with extreme skew in gender distribution.

We conclude that projection along the gender subspace (although potentially noisy for single terms) is, on average, a suitable proxy for stereotypical association with gender. 

\subsubsection{Modeling stereotype in query-document pairs}
\label{sec:map2}
Semantic memory is a specific aspect of human memory which holds general knowledge about concepts. It is regarded as a widely distributed neural network \citep{patterson2007:wk}.
Associative network structures, often referred to as schemas, are commonly used in neuroscience as models representing complex constructs that guide behavior \citep{ghosh2014:ms}. This suggests that any  acquisition of knowledge and culture resides in part in the formation of rich networks of concepts. The acquisition and articulation of stereotypes are not conceptually different: a recent line of work employs network analysis to study stereotypical associations as clusters and subclusters of concepts \citep{sayans2019:is}. 

We are interested in modeling the potential association of concepts, with a tendency to cluster along a gendered dimension. Search technologies play an important role in helping users to build links between concepts. When issuing a query, SE users are likely receptive to the formation of new links between concepts from their query and information found in ranked lists \citep{kay2015:ur}. If a document $d$ retrieved for a query $q$ (e.g.\ \texttt{nurse}) contains terms mostly aligned with the genderedness of $q$ (e.g.\ \texttt{care}, \texttt{woman}, \texttt{Mary}) it may end up reinforcing gender stereotype through an association of such concepts. 

\begin{figure}[!tbp]
  \centering
  \includegraphics[width=0.4\textwidth]{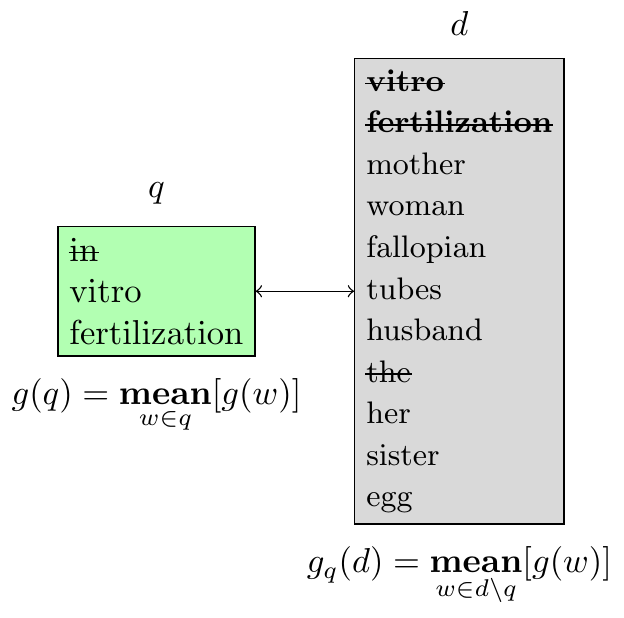}
  \caption{Concepts from query associated with concepts from document along gender dimension. Before computing the average genderedness of query and documents, stop words are removed (struck through font), and query terms which explicitly appear in a document are neglected (bold and struck through).} 
  \label{fig:map2}
\end{figure}

In order to assess the stereotypical gender agreement between $q$ and $d$, we compute their average genderedness $g(q)$ and $g_q(d)$ as schematized in Figure \ref{fig:map2}. Both queries and documents are represented as bag-of-words (stop words are removed). 
 $g(q)$ is subsequently computed as the average genderedness of remaining query terms. For $g_q(d)$, query terms are removed (bold in Figure \ref{fig:map2}) before performing the same averaging procedure. Our goal is to model the alignment of query-document concepts along stereotypically gendered lines. When computing $g_q(d)$, we therefore neglect document terms which also appear in the query, to remove the spurious bias due to redundant self-linking. For this reason, $g_q(d)$ depends on the query, as illustrated by subscript $q$. 


\subsubsection{Computing the genderedness of a ranked list}
As generally known, users seldom dabble into result pages beyond the first one, and the likelihood of view decreases with document rank \citep{jarvelin20002:cg, ferrante2014:iu,  joachims2017:ai}. Performance metrics in IR have taken this aspect into account, assigning more importance to top-ranked rather than low-ranked documents \citep{jarvelin20002:cg, moffat2008:rb}. A widely-adopted evaluation measure, based on this user model, is the Discounted Cumulative Gain \citep{jarvelin20002:cg}, which weighs documents according to a coefficient that decreases with rank in a logarithmic fashion. This weighing scheme is applied to measure the effectiveness of a ranked list based on the relevance and position of the documents within it. Our approach is identical, except for our focus on genderedness rather than relevance. 

\begin{figure}[!tbp]
  \centering
  \includegraphics[width=0.5\textwidth]{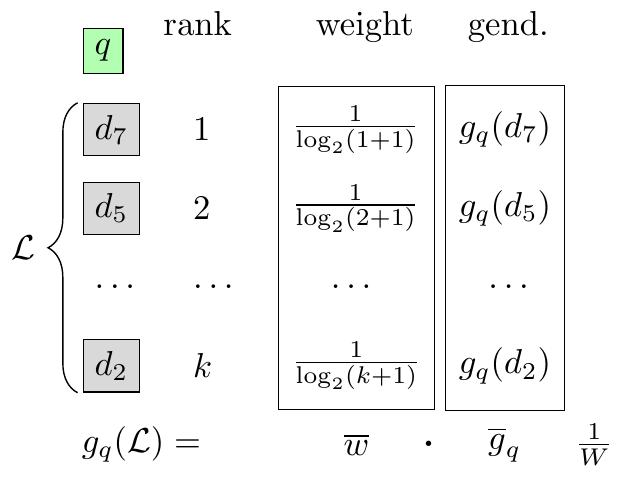}
  \caption{Genderedness of ranked list is computed as a weighted average of the genderedness of each document retrieved, with weight computed according to a rank-based logarithmic discount. Note that in the calculation of $g_q(\mathcal{L})$, without any loss of generality, we opt for a base-2 logarithm. $W$ is a normalizing constant, i.e.\ the sum of elements in $\overline{w}$}
  \label{fig:map3}
\end{figure}

Figure \ref{fig:map3} shows a ranked list $\mathcal{L}$ of documents $d_i$, retrieved for a query $q$. A vector of weights $\overline{w}$ is computed with a rank-based logarithmic discount and normalized. The genderedness of a ranked list $g_q(\mathcal{L})$ is calculated as the weighted average of the genderedness of documents in $\mathcal{L}$ with weight vector $\overline{w}$.

\begin{definition}{Genderedness of a ranked list}
\label{def:grl}

\noindent Let $\overline{w}=\left [ w_1, \dots w_K \right ]$ be a vector of weights such that $K$ is the length of the ranked list, $W = \sum_{k = 1}^{K} w_k$ and $w_k=1/ \log_2(r_k+1), k \in [1,K]$. Then, the genderedness of $\mathcal{L})$ is defined as 
  \begin{align}
  \label{eq:gsr_list}
      g_q(\mathcal{L}) = \frac{1}{W} \sum_{k=1}^K w_k \cdot g_q(d_k),
  \end{align}
\noindent with $r_k$ being the rank of document $d_k$, and $g_q(d_k)$ its genderedness. 

\end{definition}

As a toy example, which will be expanded and further discussed in Section \ref{sec:toy}, suppose we have the following setting with a single-term query and two retrieved documents:

\noindent $q=\texttt{electrician}$

\noindent $d_1=\texttt{The man is an electrician.}$

\noindent $d_2=\texttt{The woman is an electrician.}$

\noindent $\mathcal{L}=[d_1, d_2]$.

\noindent Then, according to Definition \ref{def:grl}, the genderedness of ranked list $\mathcal{L}$ is computed as follows:

\begin{align*}
    g_q(\mathcal{L})=\frac{1}{W} \left [ \frac{1}{log(2)} g_q(d_1) + \frac{1}{log(3)} g_q(d_2) \right ] = -3.8\mathrm{x}{10^{-3}},
\end{align*}

\noindent where $d_2$ is less important in this weighted average, being the last document in $\mathcal{L}$. Its genderednesss $g_q(d_2)$ is thus discounted accordingly, and the negative value of $g_q(d_1)$, albeit smaller in modulo than that of $g_q(d_2)$, ends up prevailing. 

\subsubsection{From ranked list to search history}
Multiple search results constitute a search history which may reinforce gender stereotypes. If the language of documents in ranked lists (more specifically their genderedness) consistently agrees with that of user's queries, it is reasonable to assume that the search history supports concept clustering along a gender-stereotypical dimension.

More precisely, given a set of queries $\mathcal{Q}$ and a set of ranked lists (one per query) returned by a system $s$, we compute a linear fit between query genderedness $g(q)$ and ranked list genderedness $g_q(\mathcal{L})$, considering it a summary of the GSR carried out by $s$ on $\mathcal{Q}$.

\subsection{Measurement}
\label{sec:measurement}

Below is a summary of the steps to measure GSR:

\begin{itemize}
    \item Genderedness of a word $w$ is measured as its projection along the gender direction (Equation \ref{eq:gend_score}). 
    \item Genderedness $g(q)$ of a query is defined as average genderedness of the terms in the query, after removing stop words.
    \item Genderedness  $g_q(d)$ of a document $d$ relevant for a query $q$, is computed as average genderedness of its terms, neglecting stop words and query terms.
    \item Genderedness $g_q(\mathcal{L})$ of a ranked list $\mathcal{L}$ is computed as a weighted average of documents' genderedness. An inverse logarithmic function of rank determines the weight of each document.
    \item Given a set of queries $\mathcal{Q}$, a set of ranked lists (one per query) retrieved by a system $s$ from a collection $\mathcal{D}$, and the linear fit between query genderedness and ranked list genderedness, the GSR is the slope $m_{s}(\mathcal{Q}, \mathcal{D})$ of the linear fit. 
\end{itemize}

\noindent Hence, GSR is formally defined as follows:

\begin{definition}{Gender Stereotype Reinforcement (GSR)}
\label{def:gsr2}

\noindent Let $\mathcal{Q}$ with cardinality $N$ be a set of queries, $g(q_i)$, $i \in [1, N]$ the genderedness of $q_i \in \mathcal{Q}$; let $\mathcal{D}$ be a corpus of documents and $\mathcal{L}_i$ the ranked list provided by a system $s$ for the query $q_i$ over $\mathcal{D}$. Then, GSR of $s$ on collection $(\mathcal{Q}, \mathcal{D})$ is defined as:

  \begin{align}
     \label{eq:gsr}
    m_{s}(\mathcal{Q}, \mathcal{D}) = \frac{1}{\sigma_{g(q)}^2} \frac{1}{N} \sum_{i=1}^N (g(q_i) - \mu_q)(g_{q_i}(\mathcal{L}_i) - \mu_{q,\mathcal{L}}).
  \end{align}
  
\noindent GSR weighs the  extent to which a SE responds to stereotypically gendered queries with documents containing stereotypical language with the same polarity. In the above equation, $\mu_q, \sigma_{g(q)}^2$ are query genderedness mean and variance and $\mu_{q,\mathcal{L}}$ is the average genderedness of ranked lists of documents.
\end{definition}

We chose slope instead of correlation since the latter quantifies the predictability of the genderedness of a ranked list, given that of the query. The former also captures the extent to which highly ``female'' and ``male'' queries are answered with completely different language along the gender dimension. 

\subsection{Toy example}
\label{sec:toy}

We build a toy document collection to show how GSR captures gender stereotypes.

\noindent \textbf{Setup}: from \citep{bls2019} we sample the single-word jobs with the widest gender gap (Table \ref{tab:job}). 

\begin{figure}[!tbp]
    \centering 
    \includegraphics[width=0.8\textwidth]{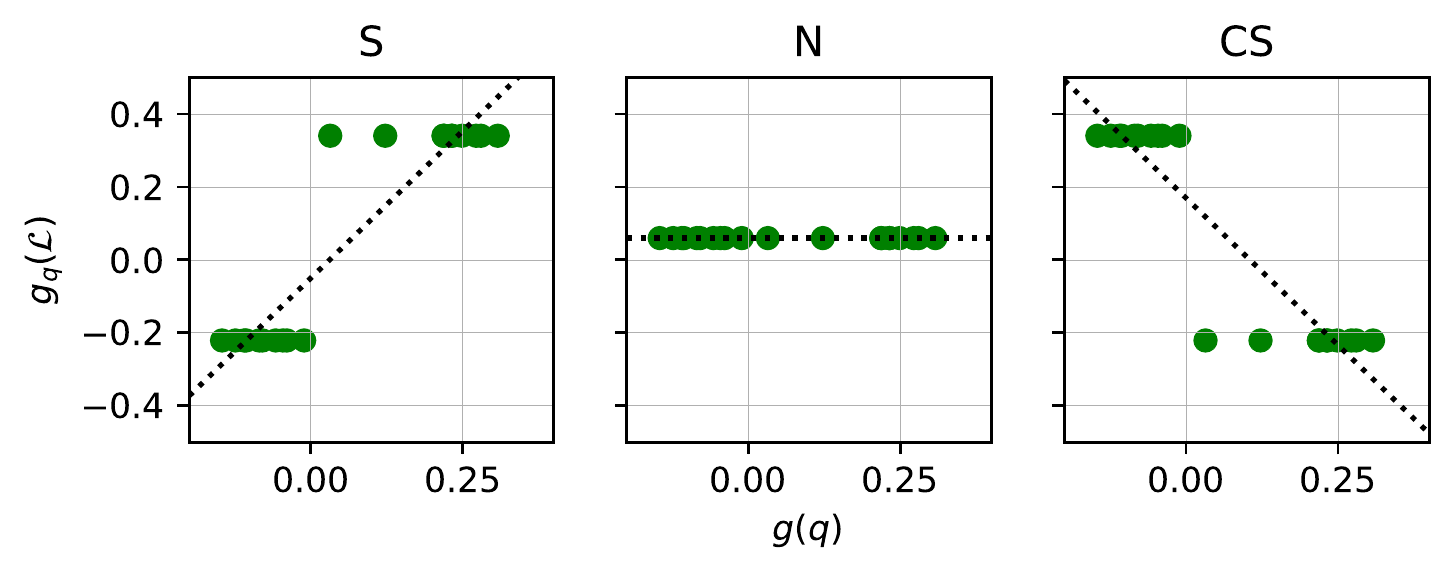}
    \caption{GSR on toy dataset $(\mathcal{Q}, \mathcal{D})$ for different retrieval systems: stereotypical (\texttt{S}), neutral (\texttt{N}),  counter-stereotypical (\texttt{CS}). GSR is the slope of linear fit, taking values $m_{\texttt{S}}(\mathcal{Q}, \mathcal{D})=1.61$, $m_{\texttt{N}}(\mathcal{Q}, \mathcal{D})=0$, $m_{\texttt{CS}}(\mathcal{Q}, \mathcal{D})=-1.61$.}
    \label{fig:toy1}
\end{figure}

\begin{itemize}
    \item $\mathcal{Q}$ is the set of (single-word) queries of occupations considered hereafter.
    
    \noindent  High female representation: hygienist, secretary, hairdresser, dietician, paralegal, receptionist, phlebotomist, maid, nurse, typist.
    
    \noindent  High male representation: stonemason, roofer, electrician, plumber, carpenter, firefighter, millwright, welder, machinist, driver.
        
    \item $\mathcal{D}$ is the set of all documents deriving from permutations of ``The \textlangle{}person\textrangle{} is a \textlangle{}job\textrangle{}'', with \textlangle{}person\textrangle{} $\in \left \lbrace \texttt{man}, \texttt{woman} \right \rbrace$ and \textlangle{}job\textrangle{} from all occupation entries in Table \ref{tab:job}.
    \item \texttt{N} is a neutral retrieval system returning, for each query, both documents (female and male) in which the query term appears (Figure \ref{fig:toy1}, center, $m_{\texttt{N}}(\mathcal{Q}, \mathcal{D})=0$).
    \item \texttt{S} is a retrieval system returning, for each query, only the stereotypical document (Figure \ref{fig:toy1}, left, $m_{\texttt{S}}(\mathcal{Q}, \mathcal{D})=1.61$). For instance, given a query about a job with a high male representation, \texttt{S} would only provide documents mentioning men.
    \item \texttt{CS} is a retrieval system returning, for each query, only the counter-stereotypical document (Figure \ref{fig:toy1}, right, $m_{\texttt{CS}}(\mathcal{Q}, \mathcal{D})=-1.61$). Contrary to \texttt{S}, given a query about a job with a high male representation, \texttt{CS} would only provide documents mentioning women.
\end{itemize}

\noindent \textbf{Discussion}: Figure \ref{fig:toy1} shows the behavior of three synthetic search engines measured by GSR. Each dot represents a query $q$, with its genderedness $g(q)$ on the $x$ axis and the genderedeness of documents retrieved for $q$  ($g_q(\mathcal{L})$) on the $y$ axis. GSR is the slope of the linear fit. \texttt{CS} and \texttt{S} are quite extreme, as they only return documents that challenge gender gap in occupations or fully reinforce it, whereas N is neutral. GSR successfully captures this aspect with zero slope for \texttt{N} and significantly non-zero slopes for \texttt{S} and \texttt{CS}, equal in magnitude and opposed in sign. The magnitude of GSR for \texttt{S} and \texttt{CS} is very large compared, for instance, against GSR values of real IR algorithms on a news collection, in the order of magnitude of $10^{-2}$ (Table \ref{tab:gsr_robust}). This depends on (1) the collection $(\mathcal{Q}, \mathcal{D})$ itself, especially conceived for \emph{direct gender stereotype}, (2) the systems \texttt{S} and \texttt{CS} which are extreme as they respond to job-related queries with documents mentioning women or men in accordance with, or in opposition to, stereotypes related to gender gaps in the occupations mentioned within a query. Overall, this experiment shows that GSR is suited to capture direct gender stereotypes. This is further  confirmed by experiments on a shared IR collection (Section \ref{sec:gsr_dir_stereo}).

\subsection{Key properties}
\label{sec:gsr_properties}

The toy example presented above defines a controlled setting where we can test the \emph{convergent validity} 
of our construct with metrics of algorithmic fairness and diversity in IR \citep{clarke2008:nd, gao2020:tc}. The IR task can be framed as a binary classification problem -- i.e.\ classifying documents as relevant or non-relevant -- with a binary protected attribute encoding whether a document is stereotypical or not. A document is deemed stereotypical for a query if it displays genderedness of same polarity. We assume a search history (in this context: solution to classification problem) to be reasonable if, for every query $q_i$, the documents included in the ranked list $\mathcal{L}_i$ contain the query term. 

Hence, in our controlled setting, for each query, such as \texttt{driver}, a maximum of two documents can be retrieved, namely \texttt{The woman is a driver} (counter-stereotypical) and \texttt{The man is a driver} (stereotypical), i.e.\ the ones which contain the term \texttt{driver} from the complete permutation described above. This is a sensible assumption and makes enumeration feasible. 

We enumerate every reasonable solution, and for each compute GSR along with the percentage of stereotypical documents among the retrieved ones, equivalent to \emph{statistical parity fairness} from \citet{gao2020:tc}, already employed in the context of fair ranking to enforce equal exposure of SE users to different topics. Results from Figure \ref{fig:conv_val} show a strong agreement between these quantities, which stems from the very definition of slope coefficient:

\begin{figure}[!tbp]
\vspace{-1.5cm}
  \centering
  \includegraphics[width=0.7\textwidth]{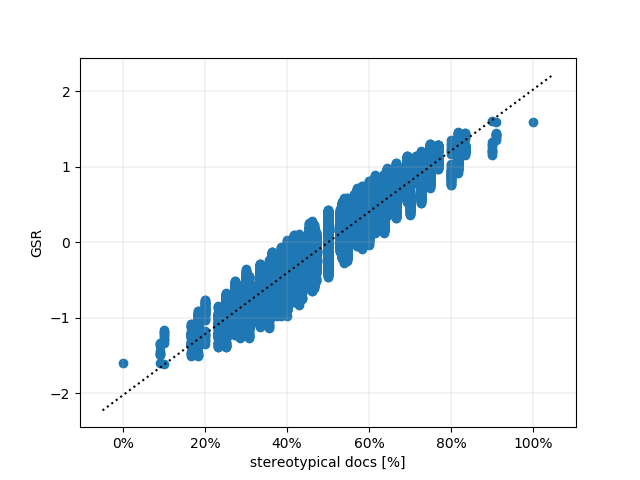}
  \caption{Agreement between GSR and percentage of stereotypical documents among retrieved ones, equivalent to \emph{statistical parity fairness} \citep{gao2020:tc}. Pearson's $r=0.92$, $p<1\mathrm{e}{-40}$.}
  \label{fig:conv_val}
\end{figure}

\begin{align}
    m_{x,y} &= \frac{1}{\sigma_x^2N} \sum_{i=1}^N(x-\mu_x)(y-\mu_y) = \frac{1}{\sigma_x^2N} \left [ \sum_{S}(x-\mu_x)(y-\mu_y) + \sum_{CS}(x-\mu_x)(y-\mu_y) \right ] ,
    \label{eq:gsr_prop1}
\end{align}

\noindent where we explicitly partitioned retrieved documents into stereotypical (S) and counter-stereotypical (CS). This partition is equivalent to topical group assignment for which statistical parity fairness enforces equal user exposure \citep{gao2020:tc}. Specializing Equation \ref{eq:gsr_prop1} for GSR, we get

\begin{align}
    m_s(\mathcal{Q}, \mathcal{D}) &= \frac{1}{\sigma^2_g(q)NW} \left [ \sum_{(q_i,d_k) \in S} \frac{(g(q_i) - \mu_q)(g_{q_i}(d_k)-\mu_{q,\mathcal{L}})}{log_2(r_k+1)} + \sum_{(q_i,d_k) \in CS} \frac{(g(q_i) - \mu_q)(g_{q_i}(d_k)-\mu_{q,\mathcal{L}})}{log_2(r_k+1)}\right ],
    \label{eq:gsr_prop2}
\end{align}

\noindent where we have used Equations \ref{eq:gsr_list} and \ref{eq:gsr}.  Documents $d_k$ which are stereotypical for a query $q_i$ (first summation in Equation \ref{eq:gsr_prop2}) bring a positive contribution to the slope coefficient $m_s(\mathcal{Q}, \mathcal{D})$, while counter-stereotypical documents bring a negative one. Equal exposure (50\% stereotypical documents) does not entail neutrality (GSR=0) and vice versa, however the two measurements are clearly correlated (Pearson's $r=0.92$, significant at $p<1\mathrm{e}{-40}$). Indeed GSR is a measure of \textit{weighted} statistical parity, between stereotypical and counter-stereotypical documents, with weight proportional to genderedness of query times genderedness of document. Although not central in this toy example, document position in ranked list $r_k$ is a further weighing factor through logarithmic discount.

A discussion focused on \emph{discriminant validity} (Section \ref{sec:construct_val}) of GSR is due. In any sensible text corpus, words from the same domain (e.g.\ medicine) and \emph{a fortiori} subdomain (e.g.\ gynecology) are likely to co-occur and their word vectors will end up close to one another, duly capturing their semantic proximity.  At the same time, in order to satisfy an information need (e.g.\ query ``in vitro fertilization''), it will be necessary to employ the specific language of relevant fields to which the query pertains (such as gynecology). For this reason, some query-document agreement shouuld be expected in language and, more specifically in genderedness. Thus, in non-trivial settings, any reasonable SE is expected to have positive GSR. Figure \ref{fig:map2} is a real example of this aspect, depicting a query and a relevant document (both represented as bags-of-words, the latter subsampled for brevity) taken from Robust04 collection \citep{harman1992:tp}. The document surely contains domain-specific language, however it is also centered around female entities, echoing old gender roles in the framing of involuntary childlessness \citep{mumtaz2013:ui}.

In other words, if a kernel of truth is present in some stereotypes \citep{penton2006:pj}, positive GSR captures a kernel of relevance, and its value is fundamentally influenced by documents available to respond to a query. Bearing such aspect int mind, it is fundamental to provide a baseline GSR for relevant documents. System N is an example of such baseline in the trivial setting of Section \ref{sec:toy}. Upward deviation of GSR from this baseline (as computed differentially or through a ratio) is regarded as the SE's contribution towards reinforcement of gender stereotype. The baseline GSR for relevant documents captures a mixture of \emph{historical bias} \citep{suresh:2019fu} and inevitable domain-specificity of language. For this reason, when measuring GSR, it is important to have a list of relevant documents as a baseline, which is to be externally validated and reasonably regarded as a ground truth.

\section{Evaluation} 
\label{sec:experiments}
\subsection*{Methodology}
GSR captures both problematic query-document associations by a SE, and domain-specificity of language embedded in a collection of documents and queries. Hence, a perfect SE, that retrieves all and only the relevant documents for each query, is expected to have positive GSR. This stems from the fact that a document is more likely to be relevant for a query if it contains specific language from the query domain - and words from the same domain tend to cluster together in the WE space, and consequently in the gender subspace. 

We can use GSR of the perfect SE as a baseline against which to compare real SEs. In other words, a SE can be said to counter gender stereotypes, even if it displays positive GSR, as long as its GSR is smaller than that of the perfect SE. Conversely, a SE which reinforces gender stereotypes will have a larger GSR. For this reason, we perform tests on shared test collections based on the Cranfield paradigm \citep{cleverdon1997:ct}, for which relevance judgements have been provided by qualified human assessors. 

Our source code and data are publicly available for reproducibility purposes.\footnote{https://github.com/alessandro-fabris/gsr}

\subsection{Synthetic dataset}

\label{sec:toy2}

In this section we introduce a synthetic example, similar in setting to the toy example presented in Section \ref{sec:toy}, exemplifying \emph{indirect} gender stereotype and GSR's ability to capture it.

\noindent \textbf{Hypothesis}: GSR can measure \emph{indirect} gender stereotypes (Definition \ref{def:igs}) stemming from clustering of concepts and language segregation along the gender direction, such as the association of stereotypically gendered occupations and traits. For example, GSR should highlight situations where SEs respond to queries about jobs with strong male representation with documents that  focus on traits related to agency.

\noindent \textbf{Setup}: We build a synthetic dataset of queries $\mathcal{Q}$ and documents $\mathcal{D}$ where a SE might promote gender stereotypes, or counter them. We simulate three SEs, designed to be stereotypical, counter-stereotypical or neutral and check whether GSR captures this aspect. Below we summarize the dataset $(\mathcal{Q}, \mathcal{D})$ and the simulated SEs working on the dataset.

\begin{itemize}
    \item $\mathcal{Q}$ is the set of (single-word) queries consisting of occupations with large gap in gender representation from table \ref{tab:job}.
    \item $\mathcal{D}$ is the set of all documents deriving from permutations of ``The \textlangle{}job\textrangle{} is \textlangle{}adjective\textrangle{}'', with \textlangle{}job\textrangle{} from Table \ref{tab:job} and \textlangle{}adjective\textrangle{} from Table \ref{tab:agency_vs_communion}. The adjectives considered are commonly used to assess gender stereotypes held by a population, and are descriptive of \emph{communion} (commonly considered a female trait) and \emph{agency} (often associated with males). 
    
    
    
    
    
    
    \item N is a neutral retrieval system (search engine) returning, for each query, all documents in which the query term appears.
    \item S is a retrieval system returning only the stereotypical document in which the query term appears. Predominantly female (male) jobs are therefore associated to communion (agency) adjectives - e.g. \texttt{The plumber is hardworking}.
    \item CS is a retrieval system returning only the counter-stereotypical document. Predominantly female (male) jobs are associated to agency (communion) adjectives.
\end{itemize}

\begin{figure}[!tbp]
    \centering
    \includegraphics[width=0.8\textwidth]{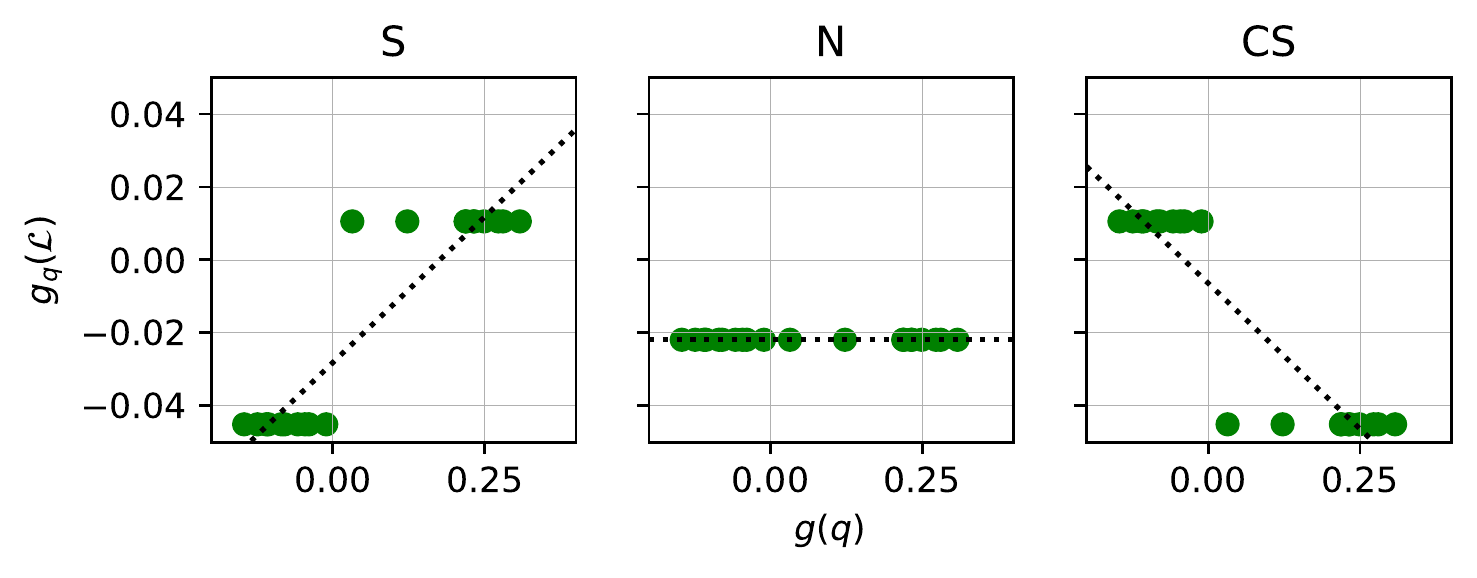}
    \caption{GSR on synthetic dataset for different retrieval systems: stereotypical (S), neutral (N),  counter-stereotypical (CS). GSR is the slope of linear fit, taking values $m_{\texttt{S}}(\mathcal{Q}, \mathcal{D})=0.16$, $m_{\texttt{N}}(\mathcal{Q}, \mathcal{D})=0$, $m_{\texttt{CS}}(\mathcal{Q}, \mathcal{D})=-0.16$.}
    \label{fig:toy2}
\end{figure}

\noindent \textbf{Discussion}: Results are summarized in Figure \ref{fig:toy2}. System S reinforces indirect stereotypes, since it links occupation and personality roles along gender-stereotypical lines, strengthening gender clusters. The proposed measure successfully captures this aspect, along with the neutral nature of N and the counter-stereotypical nature of CS. GSR can detect indirect gender stereotypes captured by underlying WEs.  

\FloatBarrier

\subsection{Real document collection}

We demonstrate our approach on a widely-used TREC evaluation  collection: TREC 2004 Robust Track \citep{harman1992:tp}, dubbed hereafter Robust04. This collection consists of about 528K news documents and 249 queries. The domain, news, is one where the relevance of web SEs is well established in mediating user access \citep{hong2018:ip}.


\subsubsection{Preliminary qualitative analysis}

\noindent \textbf{Objective}: We evaluate qualitatively whether Robust04 is an interesting collection for GSR analysis, i.e.\ whether the most gendered queries according to Word2Vec (\texttt{w2v}), contributing the most to GSR in Equation \ref{eq:gsr}, contain recognizable gender stereotypes. We have selected this candidate collection for three main reasons: (1) the large number of queries ($N=249$) for which relevance judgements from human annotators are available; (2) the domain, news, where the importance of SEs in mediating user access is wide and well-established \citep{hong2018:ip}, and (3) its relevance within the IR community.

\begin{figure}[!tbp]
  \centering
  \subfloat[Female]{\includegraphics[width=0.5\textwidth]{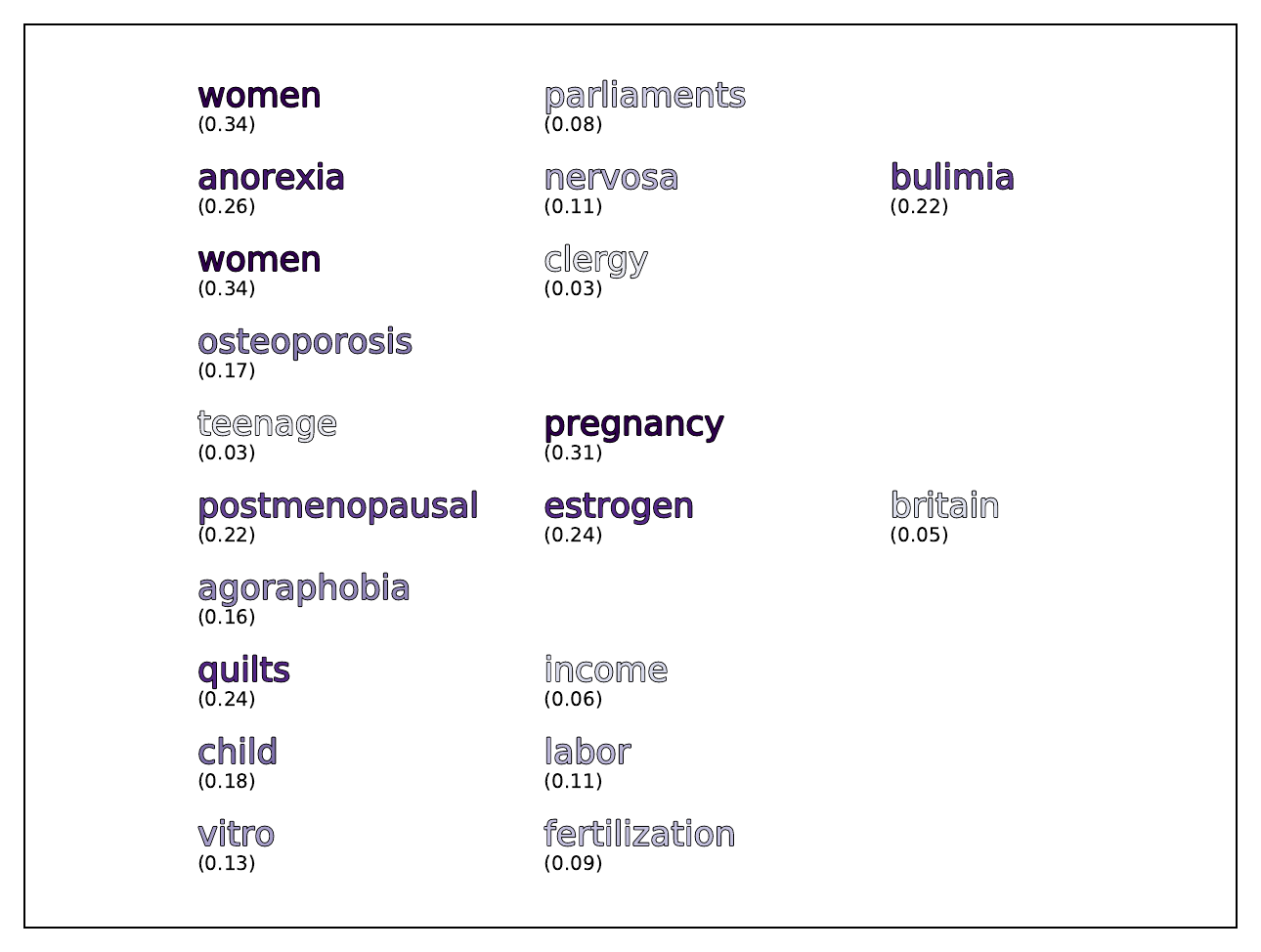}\label{fig:female_q}}
  \hfill
  \subfloat[Male]{\includegraphics[width=0.5\textwidth]{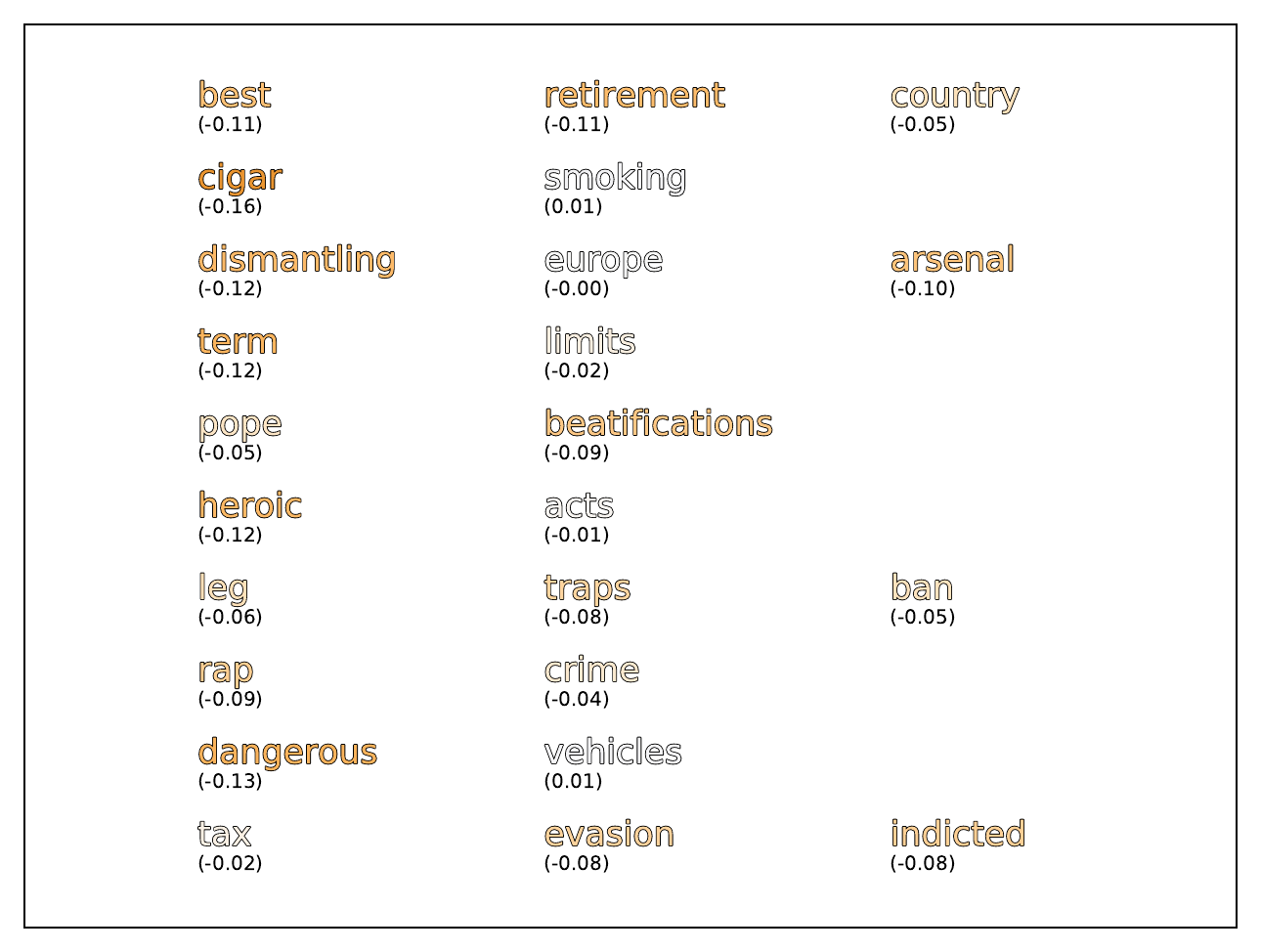}\label{fig:male_q}}
  \caption{Most gendered queries from Robust04 under \texttt{w2v}. The text is printed with color-coded gradient where strongly male words are orange, strongly female words are purple, neutral words are white. Terms' projection along gender direction can be read below each word. Stop words are removed from queries.}
\end{figure}

\noindent \textbf{Setup}: We restrict our analysis to topic titles, inspecting the 10 most ``female'' and ``male'' queries according to \texttt{w2v}. These are the queries $q_i$ whose title has the highest and lowest \emph{genderedness} $g(q_i)$, computed as the average projection of query terms onto the gender direction of \texttt{w2v} \citep{bolukbasi2016:mc}. 
The most gendered queries are depicted in Figures \ref{fig:female_q}, \ref{fig:male_q}.

\noindent \textbf{Discussion}: Among the most ``female'' queries, few are \textit{intrinsically gendered}, such as \texttt{women in parliaments} 
(topic 321) and \texttt{women clergy} (topic 445). Some more queries are \textit{biologically gendered} (such as \texttt{postmenopausal estrogen britain} and \texttt{osteoporosis} -- topics 356 and 403 respectively), describing topics biologically associated to women. The remaining queries can be described as \textit{culturally gendered}. Some are associated to disorders with apparently higher incidence on the female population (\texttt{agoraphobia} and \texttt{anorexia nervosa bulimia} -- topics 348 and 369). The final three (\texttt{quilts income}, \texttt{child labor}, \texttt{in vitro fertilization} -- topics 418, 440, 368) seem to capture unnecessary or even harmful stereotypes related to communion \citep{eagly2019:gs} and gender roles \citep{mumtaz2013:ui}. Topic 440 (\texttt{child labor}) highlights a limit of GSR and the underlying word representations. In the presence of a polysemous word, its embedding encodes a mixed representation of the different uses of said word. In this example, the embedding for \texttt{labor} has been influenced by the meaning related to giving birth, while the query has a different intent. Contextualized approaches may be useful to mitigate this issue.

All ``male'' queries seem to be  \textit{culturally gendered}, containing terms loosely related to agency (such as \texttt{dismantling}, \texttt{heroic}, \texttt{evasion}), and occupation (\texttt{retirement}, \texttt{term}), contrary to communion (\texttt{dangerous}, \texttt{arsenal}, \texttt{crime},  \texttt{traps}), and more frequently associated to men (\texttt{cigar}, \texttt{rap}). 

Overall, most of these queries associate gender (as encoded by \texttt{w2v}) with undesirable and harmful concepts. For this reason, Robust04 is a reasonable collection to study GSR.

\subsubsection{GSR on Robust04}

\noindent \textbf{Hypothesis}: SEs based on WEs, such as \texttt{w2v} and FastText (\texttt{ftt} - \citep{mikolov2018:ap}), have higher GSR than purely lexical ones, i.e.\ they are more prone to support gender stereotypes due to the problematic gender information embedded in word representations.

\noindent \textbf{Setup}: We evaluate GSR for three families of well-known IR systems, which could serve as a basis for a SE:
\begin{itemize}
    \item Lexical: these algorithms are based on matching query terms to document terms, without any information about semantics. In this group we include three models inspired by different key paradigms: the widely-used probabilistic model \texttt{BM25}~\citep{RobertsonZaragoza2009}, a popular language model~\citep{Zhai2008} (i.e.\ Language Modelling with Bayesian smoothing and a Dirichlet Prior) called \texttt{QLM}, and the classic vector space model \texttt{tf-idf}~\citep{SaltonMcGill1983}.
    \item Semantic: IR systems based on WEs have been proposed \citep{vulic2015:mc}, with the idea of exploiting the latent relationship between words encoded by the embeddings. We test \texttt{w2v\_add}, \texttt{w2v\_si} \citep{vulic2015:mc} and \texttt{ftt\_add} (\texttt{w2v\_add}'s counterpart based on \texttt{ftt} WEs) for this family. 
    \item Neural: WEs can be fed as input to neural networks,  which in turn learn to match the signals of user queries with that of relevant documents. We consider Deep Relevance Matching Model (DRMM - \citep{Guo2016}), and Match Pyramid (MP - \citep{pang2016:tm}). The embeddings used as input to these systems are \texttt{w2v} trained on Google News \citep{mikolov2018:ap}.
\end{itemize}

If a query has $K$ relevant documents, according to the assessors' judgments, we compute GSR for each system on the top-$K$ documents retrieved by it. This makes GSR of the perfect SE (retrieving all and only the $K$ relevant documents) directly comparable to that of the systems at hand.

\begin{figure}[!tbp]
  \centering
  \includegraphics[width=0.8\textwidth]{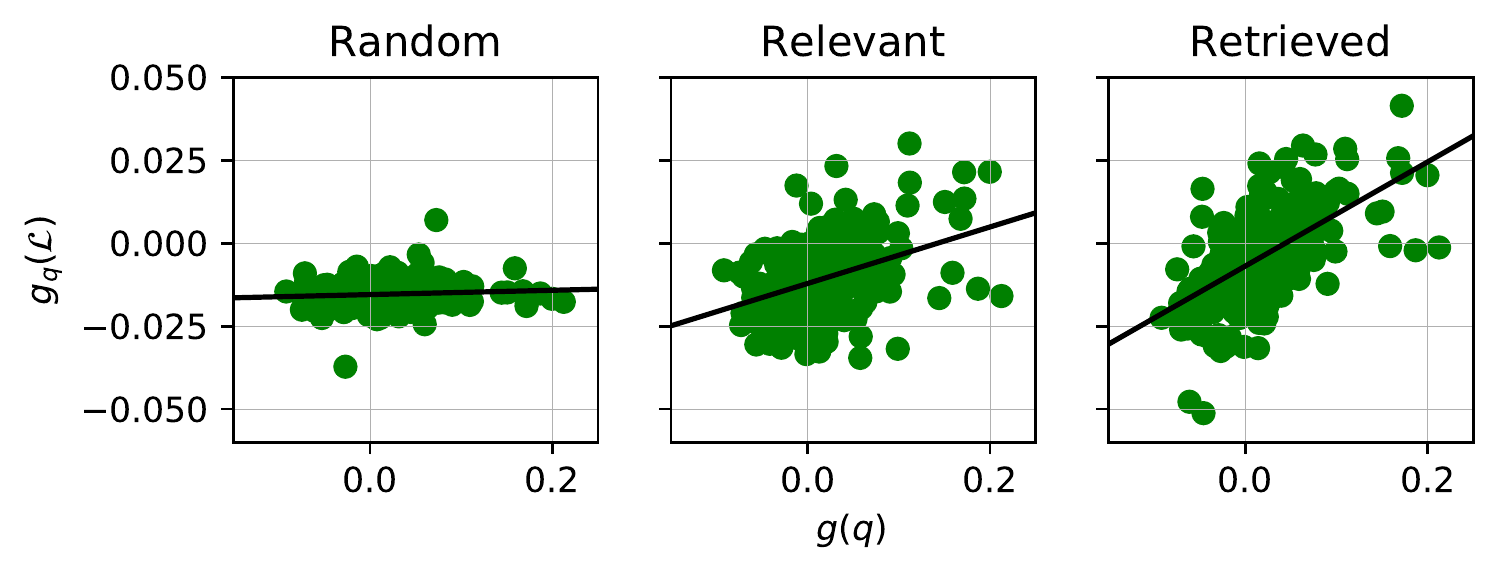}
  \caption{GSR for relevant, random and retrieved docs. The $x$ axis represents genderedness of queries $g(q)$, while $y$ axis represents the genderedness of ranked document list $g_q(\mathcal{L})$. GSR is the slope of the linear fit of the scatter plot.}
  \label{fig:gsr_rel_vs_rand_vs_ret}
\end{figure}

\noindent \textbf{Results}: As a preliminary illustration, Figure \ref{fig:gsr_rel_vs_rand_vs_ret} depicts GSR for three systems. 
\begin{itemize}
    \item On the left, a search engine retrieving random documents. 
    \item In the middle, a perfect search engine which retrieves all and only relevant documents, also ranking them perfectly according to relevance judgements.
    \item On the right, a search engine based on \texttt{w2v\_add} \citep{vulic2015:mc}.
\end{itemize}

As mentioned in Section \ref{sec:gsr_properties}, while discussing \emph{discriminant validity}, positive GSR can be associated with relevance. This is shown by the positive GSR of the perfect SE in the middle pane, compared against the near-zero GSR of random retrieval. This is not surprising; it is due to a combination of language specificity and \emph{historical bias} \citep{suresh:2019fu} potentially present in news coverage, as discussed in Section \ref{sec:gsr_properties}. For this reason, hereafter we will report for comparison the GSR of the perfect SE.

\begin{figure}[!tbp]
  \centering
  \includegraphics[width=0.9\textwidth]{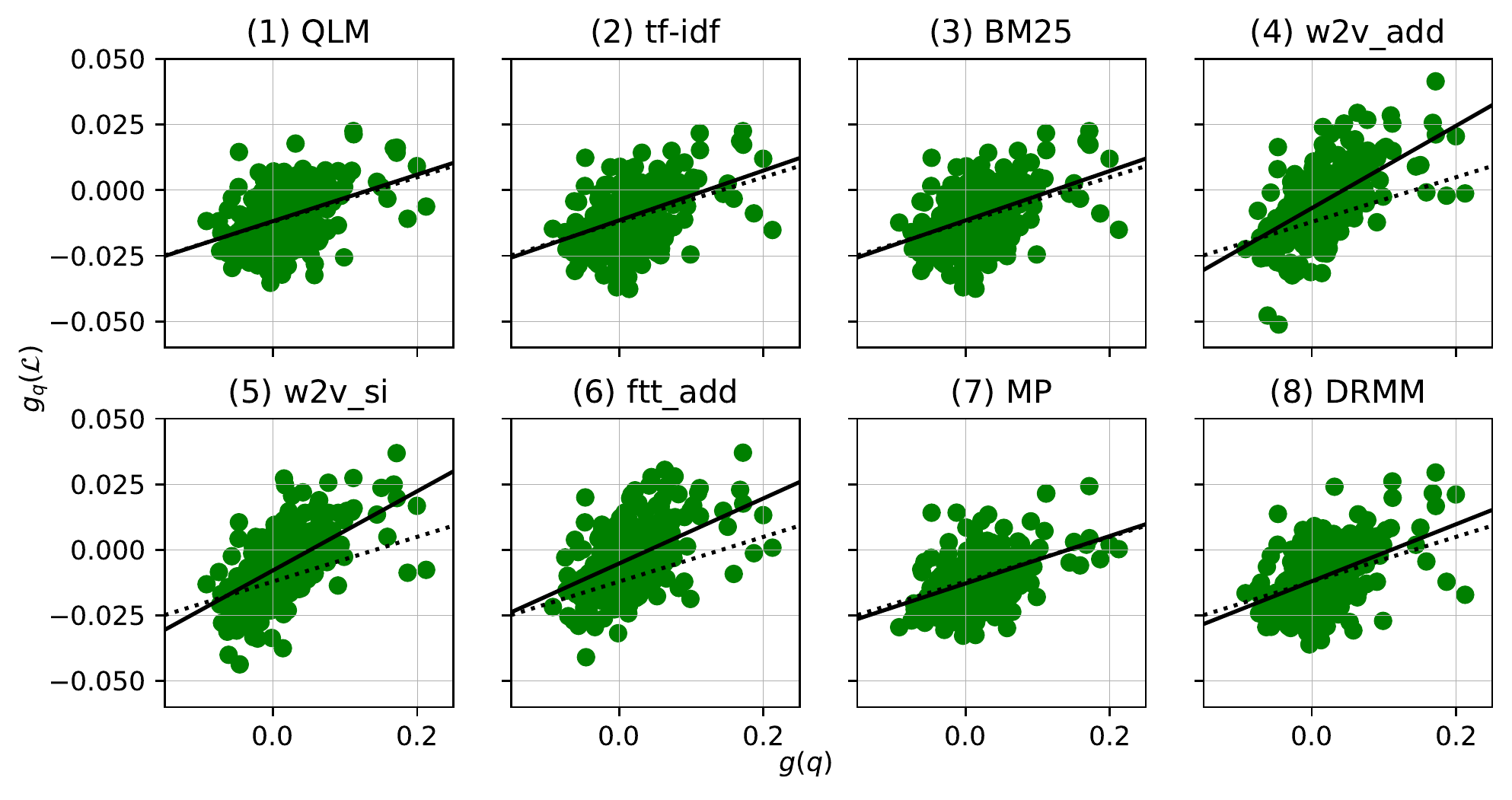}
  \caption{GSR for different systems on Robust04 according to gender direction of \texttt{w2v}. The $x$ axis represents the genderedness of queries $g(q)$, while the $y$ axis represents the genderedness of ranked document lists $g_q(\mathcal{L})$. GSR is the slope of the linear fit through the scatter plot (solid). The dashed line is the linear fit of the perfect SE, reported for comparison.}
  \label{fig:robust_gsr_summary_w2v}
\end{figure}

Figure \ref{fig:robust_gsr_summary_w2v} depicts GSR for the search results of eight different retrieval systems. Each panel contains a scatter plot of 249 different points (one for each query in Robust04), along with their linear fit (solid) and the linear fit of the perfect SE for comparison (dashed). Panels (4)-(6), depicting \texttt{w2v\_add}, \texttt{w2v\_si}, \texttt{ftt\_add}, confirm that semantic SEs have higher GSR than lexical ones, namely \texttt{QLM}, \texttt{tf-idf} and \texttt{BM25}, depicted in panels (1)-(3). This fact is easily to explain: SEs based on gender-biased WEs inherit the bias and tend to reinforce it.

Interestingly, neural systems based on the same word representation (\texttt{MP}, \texttt{DRMM}), shown in panels (7) and (8) respectively, seem to dampen this effect, thanks to successful tuning of weights during training, which reduces the importance of the (biased) gender direction in \texttt{w2v}.

As expected, semantic models based on biased WEs are likely to reinforce gender stereotypes, even when based on an IDF-inspired weighting scheme (as in the case of \texttt{w2v\_si}), aimed at assigning greater importance to terms that bear more information. On the other hand, lexical models have low GSR, comparable to that of the ideal SE.

\FloatBarrier
\subsubsection{Debiasing moderatly reduces GSR}
\label{sec:deb}
\noindent \textbf{Objective:} The gender direction along which we measure GSR can be removed from the embeddings, by means of orthogonal projection \citep{bolukbasi2016:mc}. We evaluate the impact of this operation on performance and GSR of semantic and neural SEs based on WEs.

\noindent \textbf{Setup:} For each system which relies on WEs, we repeat the previous retrieval task with three different versions of \texttt{w2v} embeddings. \emph{Regular} embeddings are the original version trained on Google News.  \emph{Debiased} embeddings are obtained by eliminating the gender direction from neutral words while maintaining it for gendered word such as \texttt{woman} \citep{bolukbasi2016:mc}. \emph{Strong debiased} embeddings  take this procedure a step further, eliminating the gender component from each word \citep{prost2019:de}. The same debiasing procedures are applied to FastText embeddings fed to \texttt{ftt\_add}.

\noindent \textbf{Discussion:} GSR values are reported in Table \ref{tab:gsr_robust}, under the header \textbf{GSR (\texttt{w2v})}. Debiasing is effective in reducing GSR for systems where it is particularly high, namely the semantic ones, purely based on WEs (\texttt{w2v\_add}, \texttt{w2v\_si}, \texttt{ftt\_add}). However, even for these systems the reduction is quite weak, ranging between $10\%$-$25\%$, as gender information leaks along different directions, orthogonal to the one that is eliminated through debiasing. This aspect has been previously studied \citep{gonen2019:lp}, to conclude that ``the gender-direction provides a way to measure the gender-association of a word but does not determine it''. Our results confirm that this is true for a measure based on the gender direction such as GSR. Furthermore, \emph{strong debiasing} brings no major advantage compared to simple debiasing.

\footnotesize
\begin{table}[tbh]
  \begin{center}
    \begin{tabular}{l  c  c  c  c}
    \toprule
      \textbf{System} & \multicolumn{2}{c}{\textbf{GSR (\texttt{w2v})}} & \multicolumn{2}{c}{\textbf{GSR (\texttt{ftt})}} \\
       & absolute & relative & absolute & relative \\
      \midrule
      perfect & $8.5\mathrm{x}{10^{-2}}$ & $0\%$ & $9.3\mathrm{x}{10^{-2}}$ & $0\%$ \\
      \texttt{w2v\_add} & & & & \\
      \hspace{0.2cm} regular & $16\mathrm{x}{10^{-2}}$ & $84\%$ & $15\mathrm{x}{10^{-2}}$ & $62\%$ \\
      \hspace{0.2cm} debiased & $14\mathrm{x}{10^{-2}}$ & $62\%$ & $14\mathrm{x}{10^{-2}}$ & $53\%$ \\
      \hspace{0.2cm} strong debiased & $14\mathrm{x}{10^{-2}}$ & $62\%$ & $14\mathrm{x}{10^{-2}}$ & $53\%$ \\
      \texttt{w2v\_si} & & & & \\
      \hspace{0.2cm} regular & $15\mathrm{x}{10^{-2}}$ & $77\%$ & $15\mathrm{x}{10^{-2}}$ & $64\%$ \\
      \hspace{0.2cm} debiased & $13\mathrm{x}{10^{-2}}$ & $55\%$ & $14\mathrm{x}{10^{-2}}$ & $53\%$ \\
      \hspace{0.2cm} strong debiased & $13\mathrm{x}{10^{-2}}$ & $54\%$ & $14\mathrm{x}{10^{-2}}$ & $53\%$ \\
      \texttt{ftt\_add} & & & & \\
      \hspace{0.2cm} regular & $12\mathrm{x}{10^{-2}}$ & $46\%$ & $15\mathrm{x}{10^{-2}}$ & $58\%$ \\
      \hspace{0.2cm} debiased & $11\mathrm{x}{10^{-2}}$ & $35\%$ & $13\mathrm{x}{10^{-2}}$ & $44\%$ \\
      \hspace{0.2cm} strong debiased & $11\mathrm{x}{10^{-2}}$ & $35\%$ & $13\mathrm{x}{10^{-2}}$ & $43\%$ \\
      \texttt{MP (w2v)} & & & & \\
      \hspace{0.2cm} regular & $9.0\mathrm{x}{10^{-2}}$ & $6\%$ & $9.6\mathrm{x}{10^{-2}}$ & $4\%$ \\
      \hspace{0.2cm} debiased & $9.0\mathrm{x}{10^{-2}}$ & $6\%$ & $9.7\mathrm{x}{10^{-2}}$ & $5\%$ \\
      \hspace{0.2cm} strong debiased & $9.0\mathrm{x}{10^{-2}}$ & $6\%$ & $9.7\mathrm{x}{10^{-2}}$ & $5\%$ \\
      \texttt{DRMM (w2v)} & & & & \\
      \hspace{0.2cm} regular & $11\mathrm{x}{10^{-2}}$ & $28\%$ & $12\mathrm{x}{10^{-2}}$ & $24\%$ \\
      \hspace{0.2cm} debiased & $11\mathrm{x}{10^{-2}}$ & $25\%$ & $11\mathrm{x}{10^{-2}}$ & $21\%$ \\
      \hspace{0.2cm} strong debiased & $11\mathrm{x}{10^{-2}}$ & $26\%$ & $11\mathrm{x}{10^{-2}}$ & $21\%$ \\
      \emph{lexical}  & & & & \\
       \hspace{0.2cm} \texttt{QLM} & $8.9\mathrm{x}{10^{-2}}$ & $4\%$ & $10\mathrm{x}{10^{-2}}$ & $12\%$ \\
       \hspace{0.2cm} \texttt{tf-idf} & $9.5\mathrm{x}{10^{-2}}$ & $11\%$ & $10\mathrm{x}{10^{-2}}$ & $14\%$ \\
       \hspace{0.2cm} \texttt{BM25} & $9.4\mathrm{x}{10^{-2}}$ & $11\%$ & $10\mathrm{x}{10^{-2}}$ & $14\%$ \\
       \bottomrule
      \end{tabular}
    \end{center}
    \caption{GSR measured according to \texttt{w2v} and \texttt{ftt} with 2 significant figures. Raw GSR values are shown (dubbed absolute), along with relative values, obtained from the former, as a percentage of the GSR value for the perfect search engine. Agreement betweeen \texttt{w2v} and \texttt{ftt}: Spearman's $\rho=0.96$,  $p<1\mathrm{e}{-10}$.}
    \label{tab:gsr_robust}
\end{table}

\normalsize
The impact on performance is very limited, as shown in Table \ref{tab:debias_performance}, reporting Mean Average Precision (MAP), precision for the top-10 ranked documents (P@10) and Normalized Discounted Cumulative Gain for the top-100  ranked documents (nDCG@100). We focus on systems based on WEs, leaving aside lexical ones, since our interest is to evaluate the impact of debiasing on classical performance measures. Our results, which are in line with prior art \citep{marchesin2019:fe}, show that debiasing (both regular and strong) produce negligible changes to average performance.  

How does such small impact on performance coexist with a significant impact on GSR (shown in Table \ref{tab:gsr_robust})? Figure \ref{fig:abs-q-gend_vs_kendall} answers this question, depicting the Kendall $\tau$ distance for ranked lists of documents (top-100) retrieved by \texttt{w2v\_add} before and after debiasing (on the $y$ axis), against the genderedness of the respective query in absolute value (on the $x$ axis). From the plot an expected property of debiasing WEs emerges in the context of IR algorithms: the most impacted queries are the ones with high genderedness, which are also the most important ones for GSR. If most queries have a low gender score, then the impact on aggregated performance will be insignificant.

\begin{table}[ht!]
  \begin{varwidth}[b]{0.48\linewidth}
    \centering
    \footnotesize
    \begin{tabular}{l  l  l  l }
      \toprule
      \textbf{System} & MAP & nDCG@100 & P@10\\
      \midrule
      \texttt{w2v\_add} & & & \\
      \hspace{0.2cm} regular & 0.067 & 0.170 & 0.174 \\
      \hspace{0.2cm} debiased & 0.068 (+2\%) & 0.171 (+0\%) & 0.176 (+1\%) \\
      \hspace{0.2cm} str. deb. & 0.068 (+1\%) & 0.171 (+0\%) & 0.175 (+0\%)  \\
      \texttt{w2v\_si} & & &  \\
      \hspace{0.2cm} regular & 0.093 & 0.213 & 0.216 \\
      \hspace{0.2cm} debiased & 0.094 $(+1\%)^\ddagger$ & 0.213 (+0\%) & 0.217 (+0\%) \\
      \hspace{0.2cm} str. deb. & 0.094 $(+1\%)^\dagger$ & 0.213 (+0\%) & 0.217 (+0\%) \\
      \texttt{ftt\_add} & & &  \\
      \hspace{0.2cm} regular & 0.056 & 0.144 & 0.150\\
      \hspace{0.2cm} debiased & 0.056 (+0\%) & 0.144 (+0\%) & 0.148 (-1\%) \\
      \hspace{0.2cm} str. deb. & 0.056 (+0\%) & 0.144 (+0\%) & 0.147 (-2\%)  \\
      \texttt{MP (w2v)} & & &  \\
      \hspace{0.2cm} regular & 0.151 & 0.283 & 0.287 \\
      \hspace{0.2cm} debiased & 0.148 (-2\%) & 0.279 (-1\%) & 0.285 (-1\%) \\
      \hspace{0.2cm} str. deb. & 0.148 (-2\%) & 0.279 (-1\%) & 0.285 (-1\%) \\
      \texttt{DRMM (w2v)} & & &  \\
      \hspace{0.2cm} regular & 0.260 & 0.423 & 0.456 \\
      \hspace{0.2cm} debiased & 0.259 $(-1\%)^\ddagger$ & 0.422 (+0\%) & 0.454 (+0\%) \\
      \hspace{0.2cm} str. deb. & 0.259 $(-0\%)^\dagger$ & 0.421 $(-1\%)^\dagger$ & 0.457 (+0\%) \\
      \bottomrule
      \end{tabular}
    \caption{Impact of regular debiasing \citep{bolukbasi2016:mc} and strong debiasing \citep{prost2019:de} on performance of models based on WEs. A Student's t test is computed between regular and debiased versions of the same algorithm, with significance at $p=0.05$ and $p=0.01$ denoted by $\dagger, \ddagger$ respectively.}
    \label{tab:debias_performance}
  \end{varwidth}%
  \hfill
  \begin{minipage}[b]{0.48\linewidth}
    \centering
    \includegraphics[width=75mm]{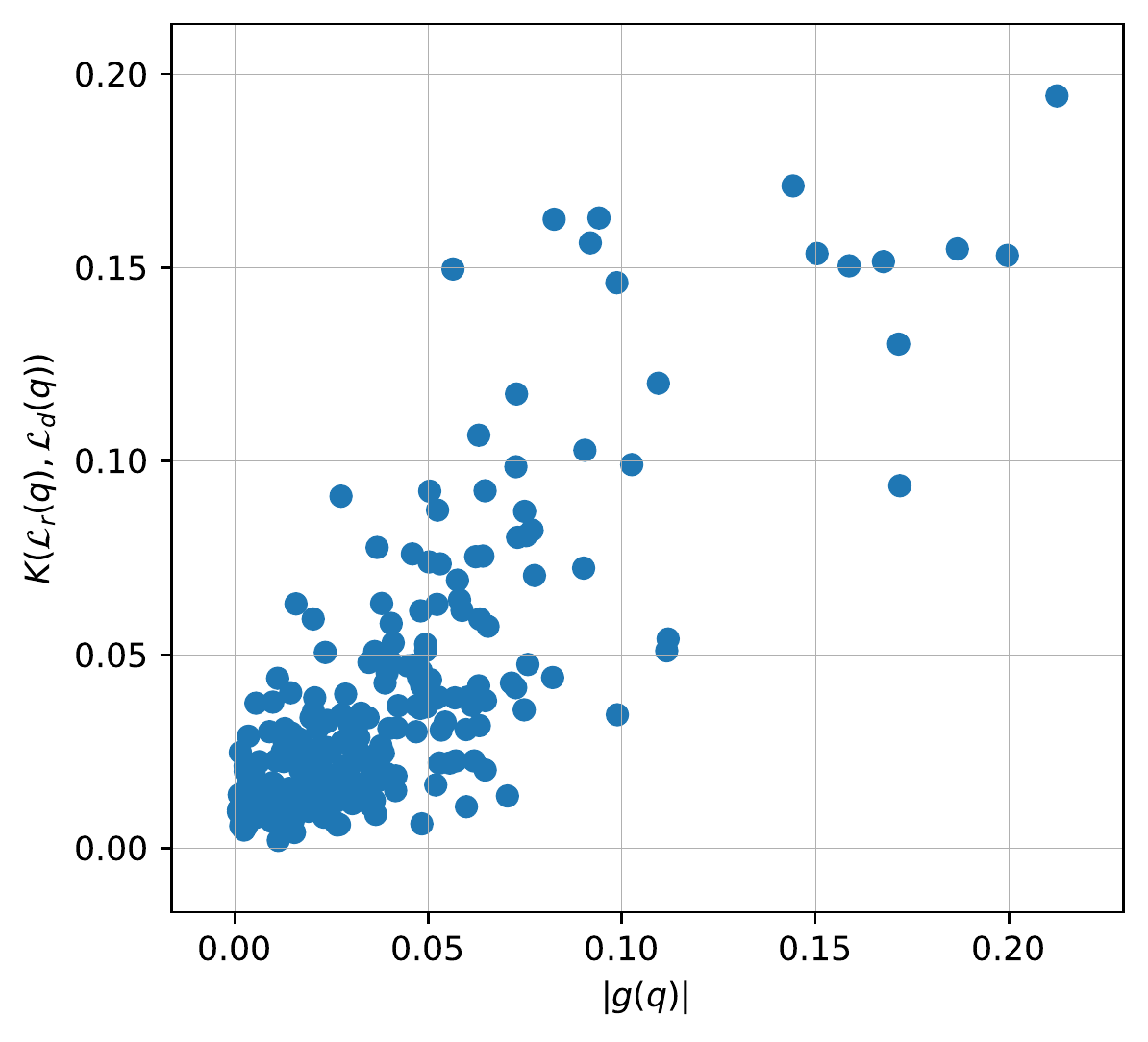}
    \captionof{figure}{Impact of regular debiasing on \texttt{w2v\_add}: absolute value of query genderedness, on the $x$ axis, and difference between top-100 documents retrieved by \texttt{w2v\_add} before and after debiasing, on the $y$ axis, as measured by Kendall $\tau$ distance. Pearson's $r=0.81$, $p<1\text{e}-50$.}
    \label{fig:abs-q-gend_vs_kendall}
  \end{minipage}
\end{table}

\subsubsection{Reliability}
\label{sec:reliability}
\noindent \textbf{Objective:} Word representations learnt with different techniques and corpora such as \texttt{w2v} (Google News) and \texttt{GloVe} (Wikipedia) have already been shown to exhibit similar bias along the gender direction \citep{bolukbasi2016:mc, caliskan2017:sd, garg2018:we}. To test the reliability of GSR, we check how dependent it is on a specific WE implementation. We do so by computing GSR based on FastText (\texttt{ftt}) WEs, trained on  Common Crawl and check its agreement with \texttt{w2v}-based GSR.

\noindent \textbf{Setup:} The procedure to isolate a gender direction and projecting word vectors from \texttt{w2v} onto it \citep{bolukbasi2016:mc} is perfectly applicable to different WEs. We compute GSR according to \texttt{ftt} embeddings, and compare it against results from previous sections obtained with \texttt{w2v}.

\begin{figure}[!tbh]
  \centering
  \includegraphics[width=0.6\textwidth]{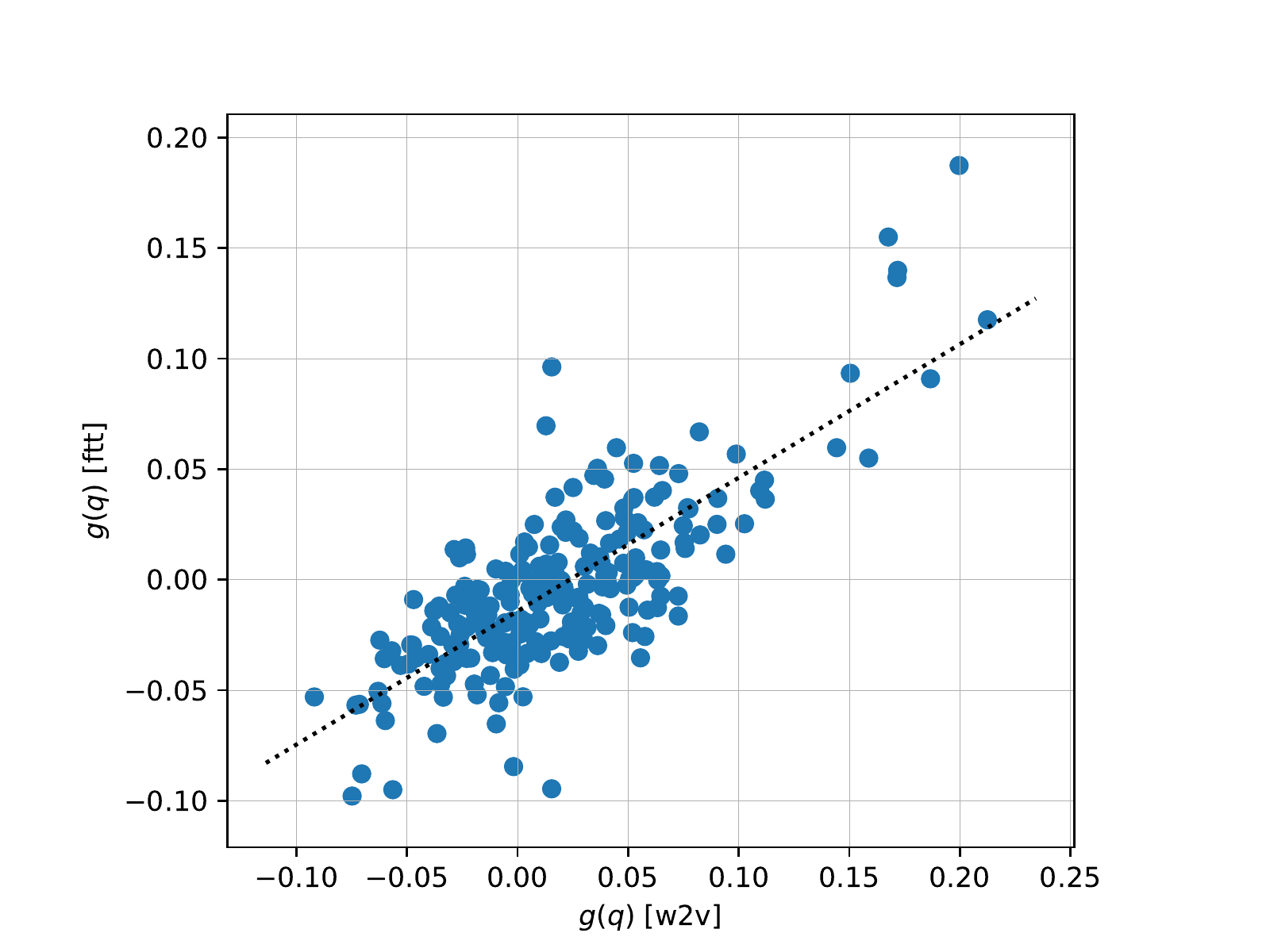}
  \caption{Genderedness of Robust04 queries, according to \texttt{w2v} ($x$ axis) and \texttt{ftt} ($y$ axis). Correlation: Pearson's $r=0.78$, $p<1\mathrm{e}{-40}$. }
  \label{fig:query_gend_w2v_ftt}
\end{figure}

\noindent \textbf{Discussion:} As a preliminary check, we compute the correlation between query genderedness measured by \texttt{w2v} and \texttt{ftt}. Figure \ref{fig:query_gend_w2v_ftt} is a scatter plot of the genderedness of 249 queries from Robust04 under \texttt{w2v} and \texttt{ftt}, which shows a strong correlation between the two (Pearson's $r=0.78$, $p<1\mathrm{e}{-40}$). This preliminary check confirms that \texttt{w2v} and \texttt{ftt} are likely to encode stereotypically gendered concepts in similar ways.

Table \ref{tab:gsr_robust} shows the values of \texttt{ftt}-based GSR, in columns 3 and 4. SEs can be ranked according to GSR scores computed with \texttt{ftt} and \texttt{w2v}. We regard the correlation of these scores as a measure of the reliability of GSR across different WEs. In other words, we would like the ranking determined by \texttt{w2v}-based GSR and \texttt{ftt}-based GSR to agree as much as possible, as measured by Spearman's rank coefficient ($\rho$). The values in Table \ref{tab:gsr_robust} (either absolute or relative) yield Spearman's $\rho=0.96$, with a $p$-value $p<1\mathrm{e}{-10}$. We conclude that GSR and the underlying gender direction is fairly reliable across \texttt{w2v} and \texttt{ftt} embeddings, despite the different text corpora from which they were learnt (Google News and Common Crawl respectively). 

We anticipated that SEs based on \texttt{w2v} (namely \texttt{w2v\_add}, \texttt{w2v\_si}) would have a higher score when GSR is measured according to the same \texttt{w2v} WEs, than when GSR is \texttt{ftt}-based. Similarly, \texttt{ftt\_add} has higher GSR when computed according to \texttt{ftt} than according to \texttt{w2v}. This is not surprising, given each SE based on word representations inherits the peculiar biases of its underlying WEs, which are the same biases that GSR captures. Despite this, \texttt{w2v}-based and \texttt{ftt}-based GSR show a solid overall agreement.

To sum up, GSR is stable across WEs that differ in architecture tweaks and choice of reasonably large training text corpus from the web domain.

\FloatBarrier

\subsubsection{GSR and direct stereotype}
\label{sec:gsr_dir_stereo}

\noindent \textbf{Hypothesis:} In order to interpret results from GSR, we investigate its relationship with explicit mentions of female and male entities. For every document, we compute a binary measure of ``intrinsic genderedness''. For the sake of simplicity, a document is considered:
\begin{itemize}
    \item Intrinsically male, if it contains more male mentions than female ones.
    \item Intrinsically female, if it contains more female mentions than male ones.
    \item Neutral, otherwise.
\end{itemize}

We hypothesize that a high GSR will result in associating stereotypically gendered queries (such as the ones depicted in Figures \ref{fig:female_q}, \ref{fig:male_q}) to intrinsically gendered documents of the same polarity. In other words, GSR should capture direct gender stereotypes (Definition \ref{def:dgs}), taking large values for SEs which associate stereotypically female (male) concepts with female (male) entities. This hypothesis relates to the \emph{content validity} of GSR, as we would expect our measure to capture this form of direct bias.

\noindent \textbf{Setup:} To assess intrinsic genderedness of documents, male and female names are sourced from \texttt{nltk}'s names corpus.\footnote{ \url{https://www.nltk.org/book/ch02.html}} Gendered nouns, adjectives and titles are obtained starting from definitional pairs and gender-specific words in Appendix C of \citet{bolukbasi2016:mc}, of which we only keep words that specifically refer to a person. Under this criterion, \texttt{aunt} is considered a female entity, whereas \texttt{pregnancy} is not. The resulting word list referring to gendered entities is reported in \ref{app:gend_ent}. 

We compare the ``perfect'' search engine (dubbed \texttt{P}) (retrieving all and only the relevant documents for each query) against one based on \texttt{w2v\_add}, which has the highest GSR among the tested systems. As a comparison, we also include \texttt{QLM} and \texttt{MP} (low GSR) and \texttt{ftt\_add} (medium GSR). Each system is compared against \texttt{P} as follows:

\begin{itemize}
    \item For each query $q$ we compute the number of intrinsically female and male documents among the ones retrieved by each SE (dubbed $f(q)$ and $m(q)$ respectively). We use their ratio as a summary of representation gap in search results ($gap(q)=\frac{m(q)}{f(q)}$). For a given list of search results $\mathcal{L}$, $gap(q)$ quantifies the extent to which documents in $\mathcal{L}$ tend to mention more male entities than female ones. 
    \item For each query $q$, we compute $gap(q)$ under \texttt{P} and \texttt{sys}, the system at hand (\texttt{w2v\_add}, \texttt{ftt\_add}, \texttt{QLM} and \texttt{MP}). 
    \item Their difference, $\Delta_{gap}(q) = gap(q)^\texttt{sys} - gap(q)^\texttt{P}$ summarizes over- or under-exposure of user to documents with male entities, compared against the ground truth of system \texttt{P}.
    \item Based on the sign of $\Delta_{gap}(q)$, we determine whether \texttt{sys} is favoring male documents (if positive) or female documents (if negative).
    \item To test our hypothesis, we compute $\text{sgn}(\Delta_{gap}(q))$ for every query and compare it against the genderedness $g(q)$ of said query.
\end{itemize} 

We expect low $g(q)$ (stereotypically male queries) to be associated with over-representation of male entities, high $g(q)$ with under-representation. Furthermore, this relationship should be strong for \texttt{w2v\_add}, weaker for \texttt{ftt\_add}, and absent from \texttt{QLM} and \texttt{MP}

\begin{figure}[!tb]
    \centering
    \subfloat[\texttt{w2v\_add}]{\includegraphics[width=0.5\textwidth]{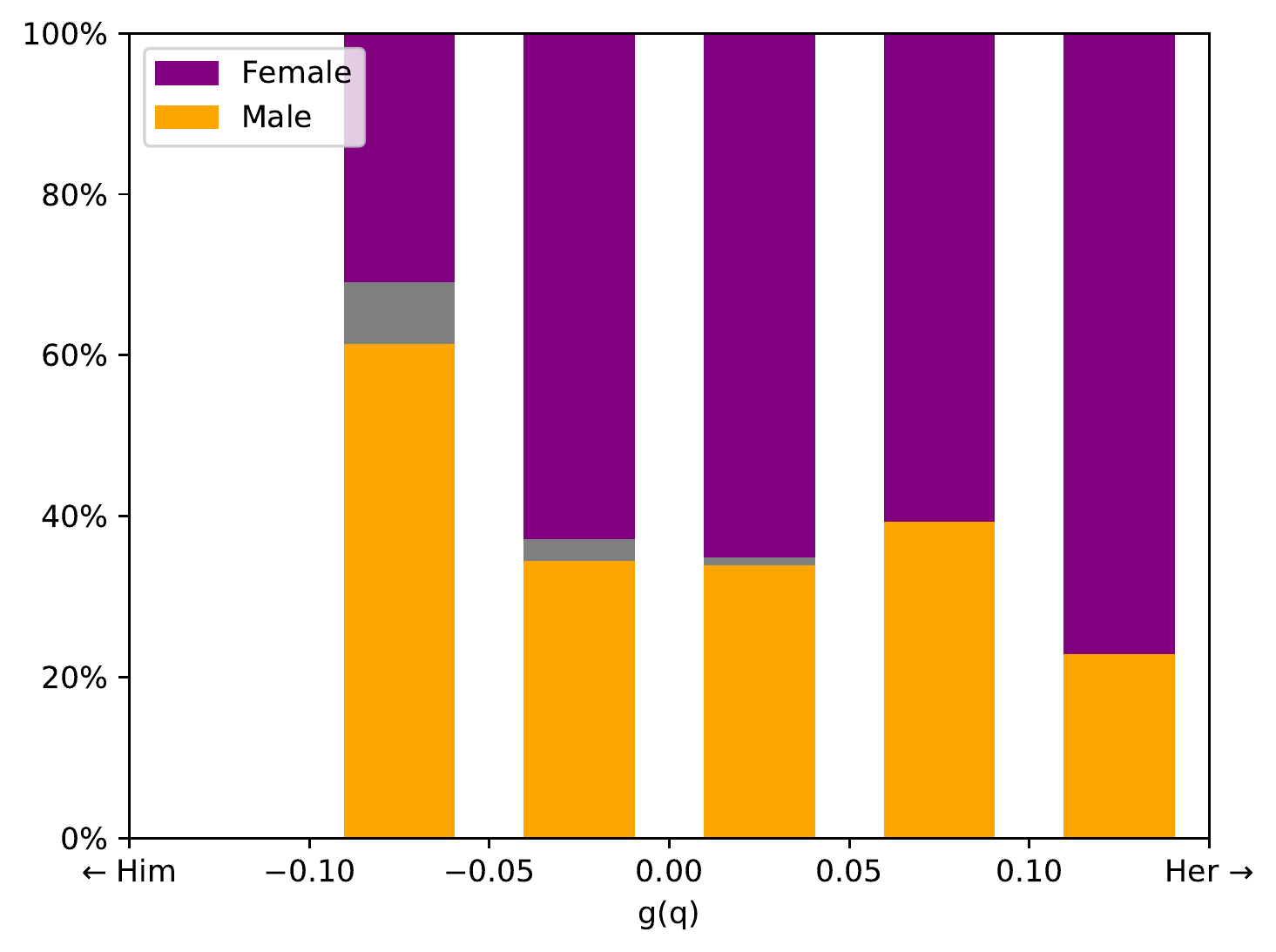}\label{fig:pv_w2v}}
  \hfill
    \subfloat[\texttt{ftt\_add}]{\includegraphics[width=0.5\textwidth]{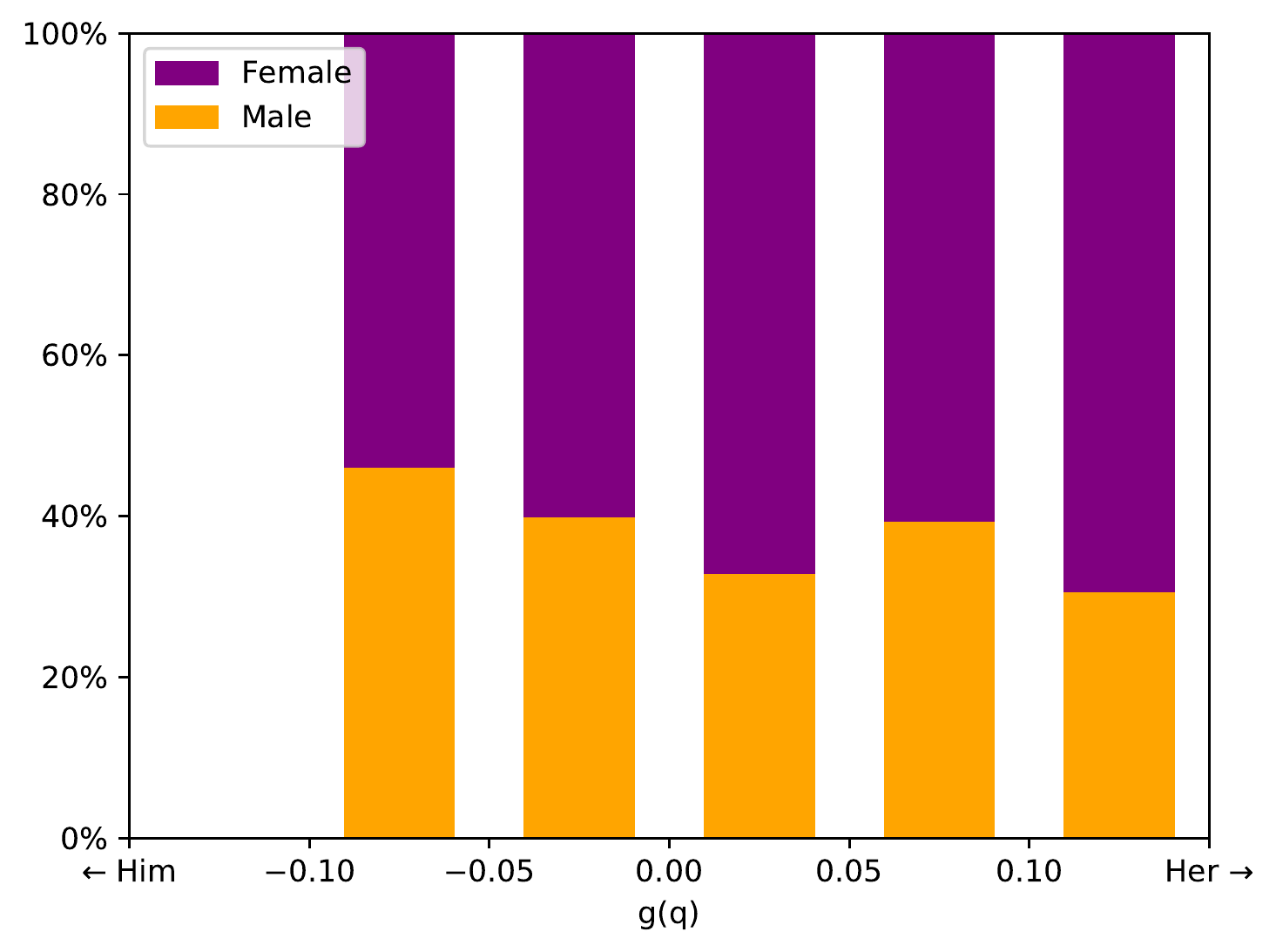}\label{fig:pv_ftt}} \\
    \subfloat[\texttt{QLM}]{\includegraphics[width=0.5\textwidth]{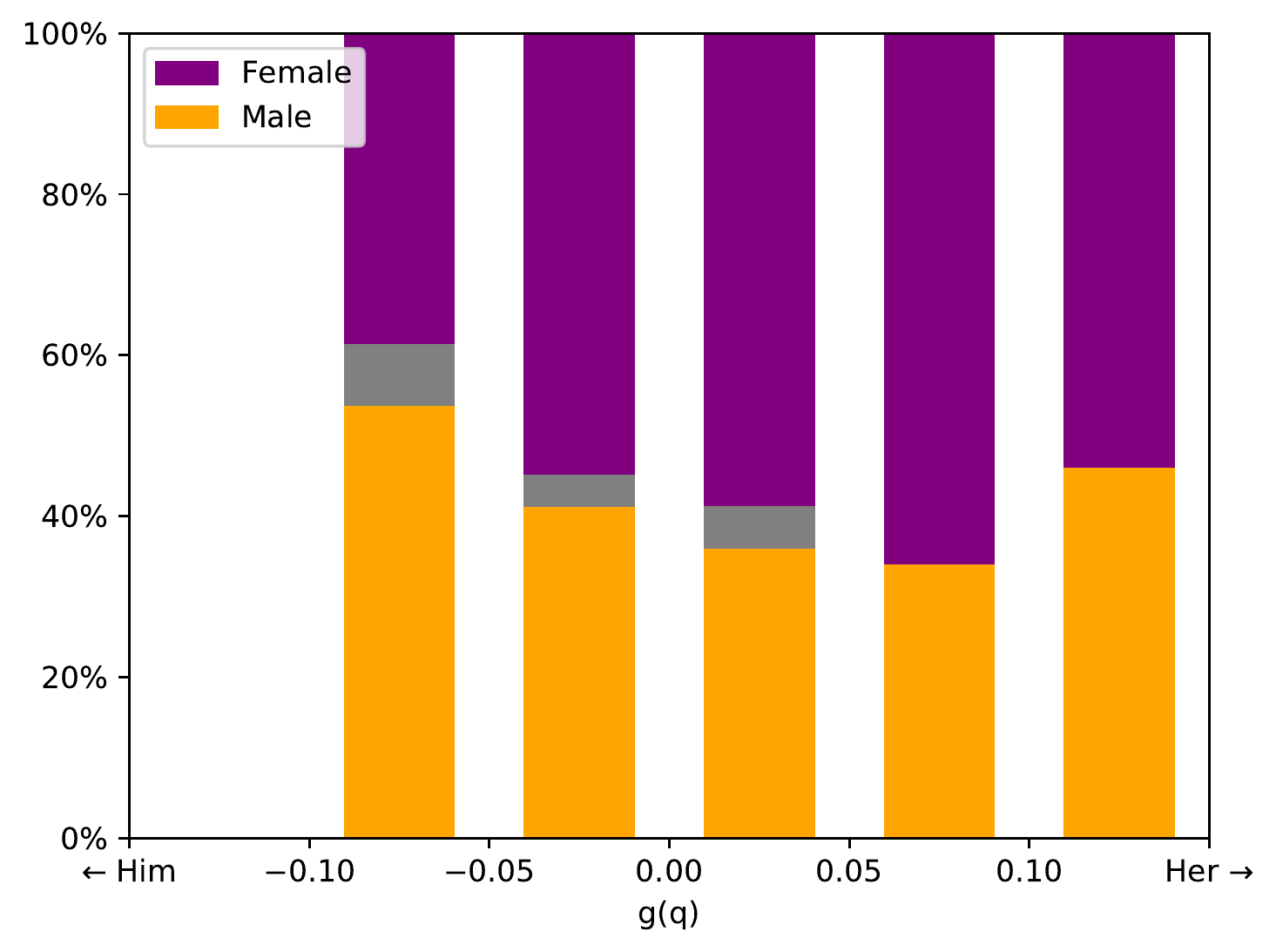}\label{fig:pv_qlm}}
  \hfill
    \subfloat[\texttt{MP}]{\includegraphics[width=0.5\textwidth]{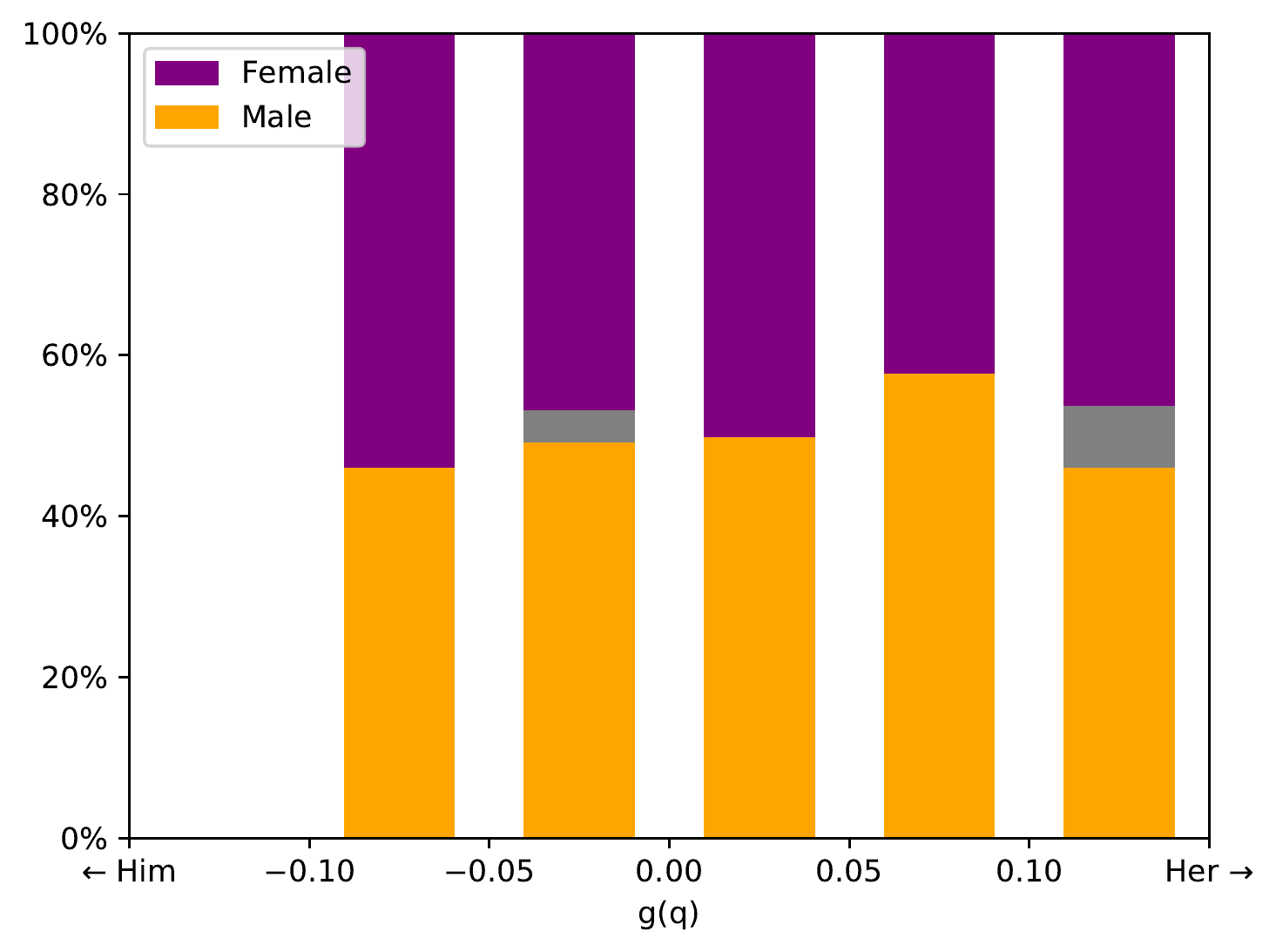}\label{fig:pv_MP}}
  \caption{On the $y$ axis, in orange (purple) percentage of queries where intrinsically male (female) documents are over-represented among retrieved ones. The complementary set, depicted in gray, is the percentage of queries for which neither gender is over-represented ($\Delta_{gap}(q)=0$). The $x$ axis is a quantization on query genderedness $g(q)$. Top panes depict semantic systems \texttt{w2v\_add} (high GSR) and \texttt{ftt\_add} (medium GSR), bottom panes show systems from the lexical and neural family (\texttt{QLM} and \texttt{MP} - low GSR). No query in Robust04 has $g(q)<-0.1$, the bin is therefore empty.}
\end{figure}

\noindent \textbf{Discussion:} Figure \ref{fig:pv_w2v} confirms our expectation for \texttt{w2v\_add}, with a clear trend along the $x$ axis. We expect said trend to be less evident for \texttt{ftt\_add}, given its lower GSR, which is confirmed by Figure \ref{fig:pv_ftt}. 

\texttt{QLM} and \texttt{MP} (low GSR) are represented in Figures \ref{fig:pv_qlm} and \ref{fig:pv_MP}. The former seems to have a weak trend similar to that of \texttt{ftt\_add}, disconfirmed however by the last bin, which contains the most gendered queries ($|g(q)|>0.1$) but does not significantly favor ``intrinsically female'' documents.  The latter displays no trend along the $x$ axis. In sum, as anticipated, no consistent trend is visible for systems with low GSR.

We conclude that GSR captures this form of direct gender stereotype: SEs with high GSR associate stereotypically gendered queries with documents mentioning people of the same gender.

\section{Conclusions and Future Work}
\label{sec:conclusion}

We defined Gender Stereotype Reinforcement (GSR) in SEs, a construct describing the tendency of a SE to reinforce direct and indirect biases about gender, which we made operational employing WEs as a measurement tool. We validated our approach against well-studied gender stereotypes from psychology literature, and exploited the framework of construct validity \citep{jacobs2019:mf} to critically evaluate our novel measure. We found that GSR captures gender stereotypes, while also being influenced by the relevance of documents retrieved for each query. This is due to the domain-specificity of language: queries and relevant documents are likely to share some specific vocabulary, whose words cluster in the embedding space and, subsequently, along the gender subspace. This aspect can be compensated when assessor judgements are available. In this regard, TREC collection Robust04 \citep{harman1992:tp} has proven to be a suitable dataset to measure the extent to which different IR algorithms reinforce gender stereotype. This is due to availability of relevance judgements, large amount of queries, interesting content of some queries from a gender stereotype perspective.

Subsequently, we studied how lexical, semantic and neural IR models reinforce gender stereotypes. We found that semantic models, based on biased WEs, are most prone to reinforcement of gender stereotypes, while neural models based on the same word representations can mitigate this effect; neural models exhibit low GSR, comparable to that of lexical systems. The reliability of these conclusions was tested with two different sets of WEs (Word2Vec and FastText), identifying strong agreement between the two measurements. 

Finally, we assessed the impact of debiasing WEs on downstream IR tasks. Regular debiasing \citep{bolukbasi2016:mc} and strong debiasing \cite{prost2019:de} have a similar effect, reducing GSR to a significant yet moderate extent. We conclude that the gender direction encoded by WEs is a useful proxy for the gender-related biases contained in the large online corpora they have been trained on. However, debiasing techniques based on projecting WEs orthogonally to the same gender direction are superficial and insufficient,  due to redundant encoding of stereotypical information. This also explains the minimum impact debiasing has on model performance.

In sum, GSR can measure associations of documents and queries along gendered lines, detecting and quantifying polarization in the language used to respond to stereotypically female and male queries. We showed that GSR captures the difference in the number of stereotypical and counter-stereotypical documents within a search history, drawing a parallel with existing statistical parity metrics \citep{gao2020:tc}.

A limitation of our measurement is the compositional model employed to assemble word scores into document scores, which does not account for syntactic structure, thus neglecting important information, such as negation.  A second drawback is the noisy nature of the gender information encoded in WEs, which should discourage the deployment of GSR on small collections, unless supported by human supervision. These observations are crucial to discuss the \emph{consequential validity} of the proposed measure. If GSR were to be integrated as part of the ranking function of a SE, it would likely favour documents which appear to be gender-neutral or counter-stereotypical for the queries issued by users. Indeed, it would be possible for providers of documents to target our measure, ensuring that their documents are not flagged as stereotypical for some queries of interest. Moreover, intrinsically gendered queries, such as \texttt{women in parliament}, require special care; low GSR may contradict user preferences.  For these reasons, we consider our operationalization of GSR  a  preliminary  attempt  to measure  gender  stereotype  reinforcement  in SEs, with limited consequential validity in fully automated contexts. Future work should include an exploration of different compositional models, based, for instance, on dependency parsers, and novel approaches to compute a gender score for words and phrases, including \emph{ad-hoc} training \citep{zhao2018:lg}. Finally, it will be interesting to measure GSR in cross-lingual scenarios; grammatical gender may pose an additional challenge in some languages, especially for the isolation of gender information along a single direction. 

To the best of our knowledge, GSR is the first measure in the domain of IR capable of quantifying a specific type of representational harm, namely gender stereotypes. This opens the possibility to quantitatively study the interplay between distributional and representational harms, which makes GSR very promising in terms of \emph{hypothesis validity} and its future uses. In the context of job search, it would be meaningful to study this interplay, due to high stakes, proven existence of biased tools 
\citep{chen2018:ii}, and availability of datasets \citep{dearteaga2019:bb}. As noticed by \citet{chen2018:ii}, search results in resume SEs, which happen to be biased with respect to gender, may lead to a dual harm: an immediate one, for the providers of CVs, competing to appear in the current search, and a long-term one, for the perception and future decisions of recruiters.

\bibliographystyle{elsarticle-harv}
\bibliography{biblio}

\newpage
\appendix
\section{Traits and terms for stereotypical associations}\label{sec:app}

\begin{table}[ht]
  \begin{varwidth}[t]{0.3\linewidth}
    \centering
    \footnotesize
    \begin{tabular}{l  l  l  l }
      \toprule
      \textbf{agency} & \textbf{communion} \\
      \midrule
      aggressive & affectionate \\
      ambitious & compassionate \\
      arrogant & emotional \\
      confident & generous  \\
      corageous &  honest \\
      critical & nurturing \\
      decisive & outgoing \\
      demanding & patient \\
      hardworking & polite \\
      independent & romantic \\
      possessive  & sensitive \\
      proud & unselfish \\
      selfish \\
      strong   \\
      stubborn \\
      \bottomrule
      \end{tabular}
    \caption{\emph{agency} vs \emph{communion}: adjectives associated to each construct \citep{eagly2019:gs}.}
    \label{tab:agency_vs_communion}
  \end{varwidth}%
  \hfill
  \begin{varwidth}[t]{0.3\linewidth}
    \centering
    \footnotesize
    \begin{tabular}{l  l  l  l }
      \toprule
      \textbf{science} & \textbf{arts} \\
      \midrule
      astronomy & art \\
      chemistry & dance \\
      Einstein & drama \\
      experiment & literature \\
      NASA & novel \\
      physics & poetry \\
      science & Shakespeare \\
      technology & symphony \\
      \bottomrule
      \end{tabular}
    \caption{\emph{science} vs \emph{arts}: associated attributes \citep{nosek2002:hi}.}
    \label{tab:science_vs_arts}
  \end{varwidth}%
  \hfill
  \begin{varwidth}[t]{0.3\linewidth}
    \centering
    \footnotesize
    \begin{tabular}{l  l  l  l }
      \toprule
      \textbf{career} & \textbf{family} \\
      \midrule
      business & children \\
      career & cousin \\
      corporation & family \\
      executive & home \\
      management &  marriage \\
      office & parents \\
      professional & relatives \\
      salary & wedding \\
      \bottomrule
      \end{tabular}
    \caption{\emph{career} vs \emph{family}: associated attributes \citep{nosek2002:mm}.}
    \label{tab:career_vs_family}
  \end{varwidth}%
\end{table}

\begin{table}[htp]
\small
  \begin{center}
    \begin{tabular}{l r r  l r r}
    \toprule
    \multicolumn{3}{c}{\large{Predominantly male}} & \multicolumn{3}{c}{\large{Predominantly female}} \\
    \textbf{occupation} & \textbf{\%F} & \textbf{\%M} & \textbf{occupation} & \textbf{\%F} & \textbf{\%M}  \\
      \midrule
      stonemason & 0.7 & 99.3 & hygienist & 96.0 & 4.0\\
      roofer & 1.9 & 98.1 & secretary & 93.2 & 6.8  \\
      electrician & 2.2 & 97.8 & hairdresser & 92.3 & 7.7\\
      plumber & 2.7 & 97.3 & dietician & 92.1 & 7.9\\
      carpenter & 2.8 & 97.2 & paralegal & 89.6 & 10.4\\
      firefighter & 3.3 & 96.7 & receptionist & 89.3 & 10.7 \\
      millwright & 5.0 & 95.0 & phlebotomist & 89.3 & 10.7 \\
      welder & 5.3 & 94.7 & maid & 89.0 & 11.0 \\
      machinist & 5.6 & 94.4 & nurse & 88.9 & 11.1 \\
      driver & 6.7 & 93.3 & typist & 86.0 & 14.0  \\
      \bottomrule
      \end{tabular}
    \end{center}
    \caption{\emph{jobs\_m} vs \emph{jobs\_f}: occupations with highest gender gap in representation \citep{bls2019}.}
    \label{tab:job}
\end{table}

\FloatBarrier

\section{Gendered entities}
\label{app:gend_ent}

\noindent The following are used in section \ref{sec:gsr_dir_stereo} to detect mentions of intrinsically gendered entities.

\noindent \textbf{Words associated with male entities:}

\begin{small}
\noindent
actor,
actors,
bachelor,
bachelors,
bloke,
blokes,
boy,
boys,
boyfriend,
boyfriends,
brother,
brothers,
brethren,
businessman,
businessmen,
chairman,
chairmen,
chap,
chaps,
congressman,
congressmen,
councilman,
councilmen,
dad,
daddy,
dads,
dude,
dudes,
ex-boyfriend,
ex-boyfriends,
exboyfriend,
exboyfriends,
father,
fathers,
fella,
fellas,
gentleman,
gentlemen,
godfather,
godfathers,
grandfather,
grandfathers,
grandpa,
grandson,
grandsons,
guy,
guys,
handyman,
handymen,
he,
him,
himself,
his,
husband,
husbands,
king,
kings,
lad,
lads,
male,
males,
man,
men,
monk,
monks,
mr,
nephew,
nephews,
pa,
prince,
princes,
salesman,
salesmen,
schoolboy,
schoolboys,
son,
sons,
spokesman,
spokesmen,
statesman,
statesmen,
stepfather,
stepfathers,
stepson,
stepsons,
uncle,
uncles,
waiter,
waiters.
\end{small}

\noindent \textbf{Words associated with female entities:}

\small{
\noindent
actress,
actresses,
aunt,
aunts,
ballerina,
ballerinas,
bride,
brides,
businesswoman,
businesswomen,
chairwoman,
chairwomen,
congresswoman,
congresswomen,
councilwoman,
councilwomen,
daughter,
daughters,
exgirlfriend,
exgirlfriends,
ex-girlfriend,
ex-girlfriends,
female,
females,
gal,
gals,
girl,
girls,
girlfriend,
girlfriends,
godmother,
godmothers,
granddaughter,
granddaughters,
grandma,
grandmas,
grandmother,
grandmothers,
her,
hers,
herself,
hostess,
hostesses,
housewife,
housewives,
lady,
ladies,
ma,
maid,
maiden,
maids,
mama,
mom,
mommy,
moms,
mother,
mothers,
ms,
mrs,
niece,
nieces,
nun,
nuns,
princess,
princesses,
queen,
queens,
schoolgirl,
schoolgirls,
she,
sister,
sisters,
spokeswoman,
spokeswomen,
stepdaughter,
stepmother,
waitress,
waitresses,
wife,
wives,
woman,
women.
}

\end{document}